\documentclass[11pt]{article}
\usepackage{graphicx}
\usepackage{comment}
\usepackage{authblk}

\usepackage{amsmath}
\usepackage{amssymb}
\usepackage{amsthm}
\usepackage{amsmath}
\usepackage{times}
\usepackage{graphicx}
\usepackage{color}
\usepackage{multirow}
\usepackage[mathscr]{euscript}
\usepackage{braket}
\usepackage{mathtools}
\usepackage{enumerate}
\usepackage{braket}

\usepackage{tikz}
\usetikzlibrary{quantikz,fit,backgrounds}
\usepackage{amsmath,amssymb}
\usepackage{graphicx}
\usepackage{subcaption}       
\usepackage{booktabs}         

\usepackage{url}
\usepackage{parskip}
\usepackage{listings}
\usepackage{xcolor}

\usepackage{hyperref}
\hypersetup{
    colorlinks=true,
    linkcolor=blue,
    filecolor=blue,      
    urlcolor=blue,
    citecolor=blue
    }

\newtheorem{theorem}{Theorem} 

\usepackage[margin=0.75in]{geometry}
\usepackage{amssymb}
\usepackage{amsmath}
\usepackage{amsthm}
\usepackage[acronym]
{glossaries}
\usepackage{nomencl}
\usepackage[font=small]{caption}
\makenomenclature

\makeglossaries

\graphicspath{ {.pdf} }
\usepackage{wrapfig}
\usepackage{subcaption}

\begin{document}

\date{}

\title{Leveraging Quantum Superposition to Infer the Dynamic Behavior of a Spatial-Temporal Neural Network Signaling Model}

\author{Gabriel A. Silva \\Department of Bioengineering and Department of Neurosciences \\Center for Engineered Natural Intelligence and Kavli Institute for Brain and Mind \authorcr University of California San Diego, La Jolla CA 92093 USA \\Email: gsilva@ucsd.edu}
\maketitle

\renewenvironment{abstract}
{\begin{quote}
\noindent \rule{\linewidth}{.5pt}\par{\bfseries \abstractname.}}
{\medskip\noindent \rule{\linewidth}{.5pt}
\end{quote}
}

\begin{abstract}
The exploration of new problem classes for quantum computation is an active area of research. In this paper, we introduce and solve a novel problem class related to dynamics on large-scale networks relevant to neurobiology and machine learning. Specifically, we ask if a network can sustain inherent dynamic activity beyond some arbitrary observation time or if the activity ceases through quiescence or saturation via an ’epileptic’-like state. We show that this class of problems can be formulated and structured to take advantage of quantum superposition and solved efficiently using a coupled workflow between the Grover and Deutsch–Jozsa quantum algorithms. To do so, we extend their functionality to address the unique requirements of how input (sub)sets into the algorithms must be mathematically structured while simultaneously constructing the inputs so that measurement outputs can be interpreted as meaningful properties of the network dynamics. This, in turn, allows us to answer the question we pose.\\
\end{abstract} 

\tableofcontents

\section{Introduction} 

Research into applications of quantum computing has historically focused on a few well-established topics. These include cryptography, data security, and the simulation of complex physical and chemical systems, including quantum mechanics itself \cite{shor1997,Feynman1982,AspuruGuzik2005,Preskill2018,Fedorov2022,AuYeung2023}. 

However, exploring new problem classes that are particularly suitable for quantum computation is an active area of research. This includes both the use of existing quantum algorithms and the development of new ones. For example, Google Quantum and X Prize recently announced a competition to promote the development of practical, real-world quantum computing algorithms and applications (\url{https://www.xprize.org/prizes/qc-apps}).

One of the unique challenges of exploring the applicability of quantum computing is that there is no known or obvious process for systematically structuring problems or questions to conform to the mathematical and algorithmic requirements of quantum computation. The structure and solution of each problem are essentially a bespoke pursuit. A well-posed problem needs to leverage quantum properties such as superposition, entanglement, and interference so that measurement outcomes are meaningful and interpretable to the question being addressed. Achieving this requires structuring problems that can demonstrate the computational advantages of quantum algorithms while also showing their relevance to practical or theoretical topics and problems of interest.

\textbf{In this paper, we introduce and solve a novel problem class related to dynamics on networks, broadly relevant to neurobiology and machine learning. We demonstrate that our solution is theoretically provably solvable more efficiently using two canonical quantum algorithms and show quantum simulation results for a small network toy example implementation of the problem. The question we address is the following: We ask if a network, consisting of nodes (neurons) whose dynamics are governed by a particular neural signaling model, can sustain inherent dynamic activity beyond some arbitrary observation time or if the activity ceases through quiescence or saturation via an 'epileptic'-like state.} In a neuroscientific context, this question informs how and when brain networks can encode and represent information given their physical structural and temporal properties. In the context of machine learning, this work relates to understanding activation patterns in artificial neural networks, which have implications for stability, efficiency, and memory. 

From a computational perspective, solving this problem classically requires brute-force evaluation of the network’s global state by assessing the internal state of each node individually, an approach that is increasingly computationally expensive and eventually prohibitive for very large networks. \textbf{We show that this problem can be formulated and structured to take advantage of quantum superposition and solved efficiently in a workflow that leverages in a connected way the Grover and Deutsch–Jozsa quantum algorithms. To achieve this, we extend their functionality so that the framework we develop conforms to the unique requirements of how input (sub)sets into the algorithms must be mathematically structured, while simultaneously constructing the inputs to ensure that measurement outputs can be interpreted as meaningful properties of the network dynamics. This, in turn, allows us to answer the question we pose.}

Our intent here is twofold. First, it demonstrates a proof-of-concept application of quantum computing to a network dynamics problem with theoretical significance in neuroscience and machine learning. Second, it provides an approach for thinking about and structuring network dynamics problems in a way that exploits the advantages of quantum computing, contributing to the broader effort of exploring quantum computing's applicability to new types of problems and questions.

The logical flow of the paper is as follows: We first introduce the network dynamic model. We then formally define the problem that we aim to answer. We then describe how the quantum algorithms can be adapted and extended to solve the problem efficiently. We then provide formal proofs of complexity advantages and numerical demonstrations of speedups for the quantum versus purely classical approach to solving the problem. We conclude by constructing and illustrating a set of chained quantum circuits and a logical workflow, which we use to numerically simulate and solve representative toy examples of the problem by implementing our theoretical approach. 

The main intellectual contribution of this work is the ability to take a neurodynamical model and reframe it in a set-theoretic manner that conforms to the mathematical and computational constraints of the quantum algorithms, while simultaneously rendering the measurement outputs as interpretable behaviors of the network dynamics.

\section{Overview of the Problem Set Up and Solution} \label{sec:problem}
\subsection{Problem Set Up}
Our goal is to address a specific question regarding the dynamics of neural networks by leveraging the uniquely quantum property of superposition. The specific problem we address is the following: \textbf{After some period of dynamic evolution, can the dynamics (i.e., activity) of a neural network sustain inherent activity beyond an arbitrary observation time $T_o$, or are the dynamics guaranteed to stop, either due to 'epileptic'-like saturation or quiescence?}

By \emph{inherent dynamic activity}, we mean the ability of the neurons (nodes) that make up the network to continue firing and maintaining the network's internal dynamics autonomously without requiring an external driving input. The network functions as a self-sustaining system where activity persists due to its internal structure and the interactions between connected nodes.

By \emph{epileptic}, we mean a state where all neurons in the network fire simultaneously at the observation time $T_o$. If this occurs, the dynamics will likely perpetuate indefinitely, maintaining a state of uniform and saturated firing beyond $T_o$. More precisely, the probability of this outcome depends on the network’s connectivity, geometry, firing frequencies, and the internal dynamic state of the neurons. We assume here that when all neurons fire simultaneously, the dynamics are likely to enter this saturated state, a condition we have previously shown theoretically and in numerical experiments \cite{buibas2011,silva2019}. 

In contrast, \emph{quiescence} refers to the complete cessation of neuronal firing across the network, resulting in no continued activity without external input or stimulation. Again, this is a condition we have reported on in prior published work \cite{buibas2011}; see also Figure 1. below.

Much of the technical setup will be in defining and describing the neural dynamic model. While the model we introduce is certainly not the only model we could have explored, we chose it because:
\begin{enumerate}
    \item It is derived from foundational physical and neurophysiological principles (spatial and temporal summation) relevant to neurobiological neurons.
    \item It is computationally tractable and builds directly on prior work, but resource-intensive and scales with network size, i.e., the number of nodes and edges.
    \item It is of relevance to real neurobiological systems. We have previously shown how individual neurons and networks of neurons (connectomes) use the model's specific theoretical and computational properties as a functional optimization principle. See section \ref{sec:priorwork} below.
    \item It is of relevance to machine learning in that we have previously shown how the model can be used to carry out a type of data (vector) embedding and encoding whereby the relationships between encoded data are inherently a component of measurable distances between vectors in the embedding space \cite{George2021b,morar2023}.
    \item Its computational outputs exhibit sufficient complexity to make the application of quantum superposition both challenging and insightful.
    \item And lastly, taken all together, the model can set the stage for follow on exploratory work as a test case and benchmark for related quantum computing problem classes and models. 
\end{enumerate}

\subsection{Overview of the Solution}
After introducing the model and describing it mathematically, we show how to map the network dynamics question into a framework suitable for quantum computation. This involves formulating the problem to exploit quantum superposition using the Grover and Deutsch-Jozsa algorithms. We will show that the output measurements from the quantum computation result in interpretable results that allow us to infer whether the network dynamics can sustain inherent activity or cease through epileptic-like saturation or quiescence. By carefully structuring the problem, we will show that a quantum algorithmic solution is computationally more efficient than purely classical methods.

The Deutsch-Jozsa algorithm is a quantum computing algorithm designed to efficiently determine whether a function behaves the same way for all its inputs, outputting either all 0's or all 1's (constant), or if the output is an equal mix of 0's and 1's (balanced). In our case, we provide the algorithm a list of 0's and 1's that encode the activation (firing) patterns representing the network's dynamics. The algorithm processes this information all at once, leveraging the quantum property of superposition. If the outputs are \textit{all} 0's or \textit{all} 1's (constant), we interpret the network's activity pattern as being either in a quiescent state, where no activity happens, or in an epileptic-like state, where everything fires at once. If the outputs are not all the same, it implies the network's dynamics are more complex, with activity in the network having the potential to continue processing information.

However, there’s a challenge: the Deutsch-Jozsa algorithm requires evaluating all possible patterns of activity in the network at once, including patterns that did not actually occur in the network. This would make the algorithm’s output impossible to interpret in terms of the network's real dynamics. To work around this, we divide the problem into two parts. Critically, we use Grover's algorithm to identify and amplify \textit{only} the patterns of activity that actually occurred in the network. For the patterns that did not occur, we assign them fixed, known values (either 0 or 1), depending on how we are running the algorithm.

This approach allows us to run the Deutsch-Jozsa algorithm twice in the following way. In the first run, we assign all non-occurring patterns a value of 0. In the second run, we assign them a value of 1. The patterns of activity that actually occurred are already encoded in the collection of observed or measured values of 0's and 1's. If the output of the algorithm in the first run (where non-occurring patterns were assigned 0's) is all 1's, this implies that the activity patterns in the network were also all 0's, meaning the network was in a quiescent state. (Note that the algorithm \lq flips' the all 0 input to all 1 outputs due to technical reasons of how it operates and computes.) Similarly, if the output of the second run (where non-occurring patterns were assigned 1's) is all 0's, it implies that the network was in an epileptic-like state. If the outputs are not constant (i.e., not all 0's or not all 1's) in both cases, it implies that the activity patterns in the network were a mix of 0's and 1's, and we interpret this as the network having the potential to sustain dynamic activity. This is a \textit{unique} non-standard application of Deutsch-Jozsa that leverages the problem's structure and setup (see Section \ref{sec:cric} immediately below).
 
By interpreting the outputs of these two runs, we can infer the state of the activity pattern of the network and determine if the activity can continue or stop. \textit{The quantum computational advantage lies in the fact that we do not need to evaluate each pattern individually, one at a time. Instead, the combined use of Grover, linked with two runs of Deutsch-Jozsa, evaluates the entire set of patterns simultaneously in superposition, saving an increasing amount of computational effort as the network size increases.} The larger the network being interrogated, the greater the computational speedup and resource savings this approach achieves.

\subsection{Beyond classical oracle limitations: Why the standard criticisms do not apply} \label{sec:cric}
In this section, we attempt to address upfront how structured Grover amplification supports the practical use of Deutsch-Jozsa, given the specific construction of the problem we introduce, and why the standard criticisms and limitations of Deutsch-Jozsa in practical settings do not apply. Our goal here is to convince readers, or at least intrigue them enough, that the typical arguments against any Deutsch-Jozsa quantum advantage in a real-world problem are not valid, in the hope that they will continue reading. We assert that, in at least some instances, Deutsch-Jozsa, when combined with Grover and possibly other algorithms as part of a broader solution workflow, does confer a computational speedup and advantage. 

In the standard textbook case, Deutsch-Jozsa is analyzed under a very specific and constrained set of conditions. One is given a black-box oracle function that evaluates to the binary output
\begin{equation} \nonumber
    f:\{0,1\}^n \rightarrow \{0,1\}
\end{equation}
Built into this construction is an explicit promise that the oracle function is either constant, i.e., all outputs are either $0$ or $1$, or balanced, i.e., exactly half the outputs are $0$ and exactly half are $1$. The objective is to determine with certainty and no error whether $f$ is constant or balanced. Under these conditions, Deutsch-Jozsa is guaranteed to arrive at a solution in one attempt.

In most real-world conditions, the strict textbook promise of a function being exactly constant or exactly balanced does not apply. It simply is not a condition of relevance to the structure of real-world systems and the physical questions asked of them. Functions may be almost balanced or almost constant, but noise and approximation conditions render the exactness constraint unrealistic. The problem is that if one relaxes this constraint and allows even a small error in the acceptable evaluation of the function, then a classical randomized algorithm operating on a small subset of samples will quickly be able to distinguish between almost balanced or almost constant conditions with high probability every time. There is no advantage to a quantum algorithm like Deutsch-Jozsa, and therefore, no need for quantum computing.

However, how we structure the question we ask in this paper, and the approach we use to solve it, deviate from the textbook usage of the Deutsch-Jozsa algorithm while respecting the strict exactness constraints when we do make use of it. We do not rely on Deutsch-Jozsa to directly answer the question about network dynamics; we use it strictly as a final 'checking' substep in the workflow on a simple function that meets the textbook criteria, but which is interpretable to us in the broader context of how the problem was intentionally set up. 

Specifically, we are not evaluating a random function with noise or deviations from the constant or balanced exactness constraints. The function we evaluate is an actualized set of structured spiking events that, by construction, is (or is not) always exactly constant. We use Grover's algorithm first to efficiently identify and amplify only the subset of states that were actually realized in a specific evolution of the network dynamics, using an explicitly defined thresholding function. All non-actualized states are uniformly set to either 0's or 1's (in two different back-to-back runs). The Grover quantum pre-processing step is critical, and classical random sampling cannot provably be as efficient. This is a source of (quadratic) quantum speedup in the workflow. 

Because we construct the overall function ourselves, we can ensure an exactly constant scenario, i.e., \textit{all} states are 0 (if the network is quiescent) or 1 (if the network is epileptic). If both runs deviate from exact consistency, it can only be due to a mixture of 0's and 1's, which we can neurophysiologically interpret as a potential for continued dynamics. In other words, we guarantee the exactness constraint by construction, and can 'check' it provably faster using canonical quantum algorithms versus classically necessarily 'sequentially' checking the state of each node (neuron) in the network. 

In this workflow, Deutsch-Jozsa is merely a final validation step after marking states using Grover's algorithm. It operates on a constructively \textit{exact} constant function. But it is only a step in a sequence of steps that allow us to \textit{interpret} the final outputs of the workflow in a meaningful neurophysiological context. What we do not do is attempt to use Deutsch-Jozsa on an approximate function that relaxes the exactness constraints as the end-all and be-all.

\section{Refractory and Signal Latency Dynamics in Geometric Neural Networks}
The network dynamic model we will use operates on structural geometric networks. From a technical perspective, it is effectively a geometric extension of classical neuronal models. This model takes into account the neurobiologically canonical processes of spatial and temporal summation of discrete signaling events (action potentials) incident on nodes (neurons) within a geometric network. We define a \emph{geometric network} as a physically constructible structural network where the edges between neurons are convoluted paths with physical distances. In a neurobiological context, this corresponds to the path lengths of axons and dendrites that connect chemical synapses between neurons, making the model anatomically accurate. The inclusion of network geometry and finite signaling speeds (conduction velocities) of action potentials naturally introduces signaling latencies.

The dynamic and diverse arrival and summation of action potentials at individual neurons, combined with a neuronal refractory period (described below), determine when neurons fire. It is the interplay between these parameters that produces the resulting rich and complex dynamics at the network scale. The dynamics are sufficiently complex that it is not currently possible to theoretically predict the temporally evolving behavior of the network. Consequently, determining at an arbitrary time $T_o$ whether the dynamics will saturate (become epileptic) or quiescent requires observation of all nodes in the network. 

In the remainder of this section, we provide a self-contained technical description of the model. Readers interested in its full development, associated proofs, and theoretical and experimental results are referred to the references cited below.

\subsection{The Refraction Ratio and its Effects on Dynamics} \label{sec:refraceg}
We first motivate the model and the network dynamic problem by presenting a simple but powerful example of how network geometry and signal latencies impact dynamics when coupled with the internal processing time of signals by individual nodes, manifested as a refractory period.

A critical theoretical result of our work is the derivation of what we term the \emph{refraction ratio}. We have demonstrated mathematically (via formal proofs) that the refraction ratio imposes physical constraints on the optimal conditions for dynamic activity in networks with a physical structural geometry \cite{silva2019}. Brain networks fall into this category and are subject to these constraints. Specifically, the ratio between the distances signals must travel within the convoluted geometry of the network, and the speed at which they propagate must be carefully balanced with the time individual nodes require to process incoming information. This balance ensures coherent network dynamics. This concept is independent of scale, so it applies equally to networks of connected neurons in a particular circuit in a part of the brain as it does to connected brain regions at the scale of the whole brain. The refraction ratio encapsulates this relationship by balancing
\begin{enumerate}[i.]
    \item \textit{Local constraints:} The time each node requires to process information internally.
    \item \textit{Global constraints:} The time required for signals to traverse the network as a function of its geometry and signaling speeds (conduction velocities).
\end{enumerate}

An example of the effect of deviations from an optimized refraction ratio is shown in Figure 1. This example involves a simulated neurobiological network of 100 neurons modeled as Izhikevich neurons~\cite{Iz2003}, a simplified mathematical representation of Hodgkin-Huxley neurons. Raster plots, commonly used by neuroscientists to visualize neural activity, illustrate the dynamics of the entire network. The vertical axis enumerates each neuron from 1 to 100, while the horizontal axis represents time, in this example spanning 4 seconds (4000 milliseconds). A tick mark along the time axis for a given neuron indicates that the neuron fired (an action potential) at that time.

This network is geometric, meaning the physical distance between neurons influences dynamics (Figure 1A and B). Specifically, signal propagation speeds — action potential conduction velocities — over these distances introduce latencies (or delays). Additionally, each neuron has a refractory period, during which it cannot respond to other incoming signals.

In this example, the network was externally stimulated for the first 500 milliseconds. We then assessed its ability to sustain inherent recurrent signaling without additional external stimulation. The network exhibited recurrent, low-frequency, periodic sustained activity at the lowest signaling speed (Figure 1C top panel). However, when the signaling speed was increased by a factor of 100, no activity persisted beyond the externally driven stimulus period; all activity ceased (Figure 1C bottom panel). 

This cessation was the result of a mismatch in the refraction ratio. When signals arrived too quickly, neurons could not recover from their refractory periods. Consequently, incoming signals failed to induce downstream activations, and the activity in the network collapsed entirely. 

\begin{figure} \label{fig1}
  \includegraphics[width=\linewidth]{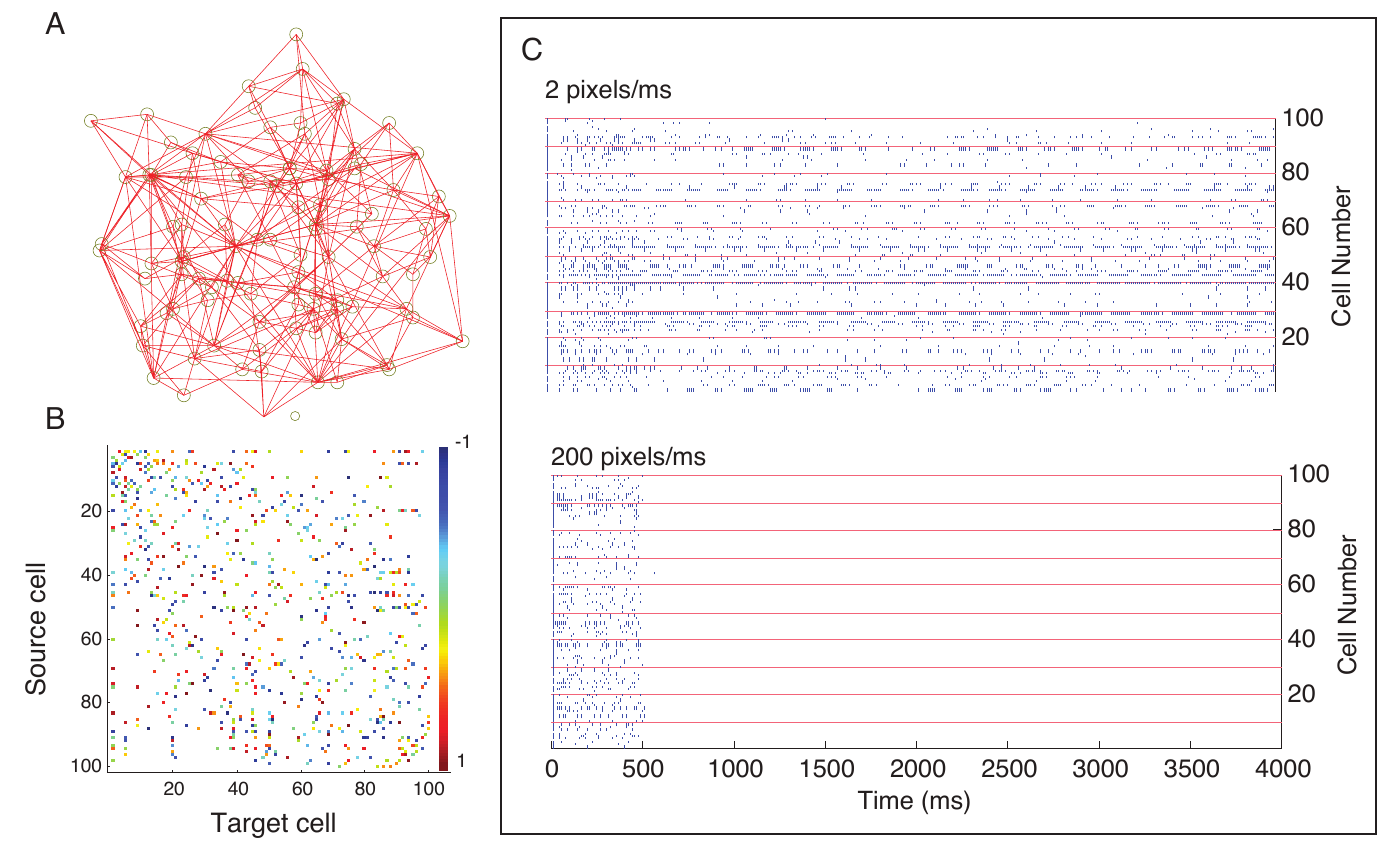}
  \caption{\textbf{Effects of signaling speed (conduction velocities) and resultant latencies on the dynamics of a geometric neural network with neuronal (node) refractory periods. A.} Graphical representation of the spatial locations of nodes and their physical (structural) connectivity in the plane. \textbf{B.} Functional connections were assigned random weights univormly between -1 and 1 on each phyical edge. An Izhikevitch model of bursting neurons was used to model the individual vertex dynamics. \textbf{C.} When the conduction velocities of signal propagation's was varied, the signaling delay distribution (temporal latencies) scaled and had a significant impact on the spike train dynamics (see text). Adapted from \cite{buibas2011}. }
  \label{fig:figure1}
\end{figure}

\subsection{Prior Work} \label{sec:priorwork}
The main theoretical results, including proofs and derivations, can be found in \cite{silva2019}. These results build on earlier findings that extended the modeling of biological geometric neural networks and their dynamics \cite{buibas2011}.

We have demonstrated that at least one class of neurons — basket cell inhibitory neurons in the cortex — has evolved specific morphological features to optimize the refraction ratio as an optimization principle, as a result enhancing signaling efficiency \cite{puppo2018}. Additionally, the neurobiological connectome of the worm \emph{C. elegans} leverages the refraction ratio and refractory dynamics to optimize signaling and coordinate motor outputs \cite{george2021}. This model has also been adapted for use in a novel machine learning data embedding method \cite{George2021b,morar2023}. 

Our most recent (unpublished) data, collected using magnetoencephalography (MEG) in humans, further suggests that a form of the refraction ratio preserves perceptual symmetries during intra- versus inter-hemispheric visual tasks. These findings underscore the model's broad applicability and relevance to biological and artificial neural network systems.

In the remainder of this section, we discuss the mathematical construction of the model and the refraction ratio, focusing on how fractional signaling summations produce activation event outputs. This aspect directly relates to the problem we address in this paper and its mapping to the quantum domain. However, we do not explore how the refraction ratio produces theoretical bounds on efficient network signaling, which is less important to the work in this paper. We also do not provide proofs for the stated theorems. Interested readers are referred to the cited references for a more comprehensive treatment.

\subsection{Mathematical Introduction to the Model} \label{sec:model}
The model describes how temporal latencies create offsets in the timing of summations for incoming discrete signaling events, such as neuronal action potentials, and how these interactions lead to the activation of downstream nodes. It captures how the timing of different signals compete to activate the nodes they arrive at. Here, \emph{activation} refers to the appropriate reaction of a node to incoming signals. For biological neurons, this usually means generating an action potential that propagates down the axon and its axonal arborizations to downstream synapses. This is followed by a period of refractoriness, during which the neuron cannot respond to new incoming action potentials. 

In a more abstract sense, the concept of activation generalizes to the production of discrete signaling events appropriate to the network (e.g., excitatory versus inhibitory), and refractoriness generalizes to the information processing time of a node required for its internal computations to result in an actionable computational output. The interplay between the temporal latencies of propagating discrete events across the network and the internal dynamics of individual nodes profoundly influences the overall network behavior, as demonstrated above in section \ref{sec:refraceg}. 

In a geometric network, temporal latencies arise from the interplay between the signaling speeds and the geometry of the network’s edges, i.e., the path lengths between connected nodes. The model assumes that each node has an associated \emph{refractory period} or \emph{refractory state}, which reflects its internal state after activation by an upstream node (or set of nodes). The node cannot respond to other incoming signals during this refractory period. We make no assumptions about the internal model producing this refractory state, which could include an internal processing period during which the node decides how to react or could be due to the physical makeup and operation of the node, or both. For example, in biological neurons, the biophysics associated with the molecular resetting of sodium ion channel inactivation domains determines the absolute refractory period. 

\subsubsection{Geometric network construction} \label{sec:geometric}
The model considers a spatial-temporal network where signals propagate between nodes along directed edges at a finite speed or conduction velocity, resulting in temporal delays (latencies) in signal arrival. These latencies naturally map a geometric representation of the network, where each node can be assigned a spatial position in $\mathbb{R}^3$ for every vertex $v_i$, with $i = 1, \ldots, N$, where $N$ is the number of nodes in the network, or its size. We use the terms \emph{vertices} and \emph{nodes} interchangeably.

Edges between nodes can follow convoluted paths, reflecting the physical geometry of connections. A signaling latency $\tau_{ij}$ is defined as the ratio between the distance traveled on an edge, $d_{ij} = |e_{ij}|$, and the speed of the propagating signal, $s_{ij}$:
\[
\tau_{ij} = \frac{d_{ij}}{s_{ij}}.
\]

The set of all edges in the network graph $G = (V, E)$ is denoted by $E = \{e_{ij}\}$, where $e_{ij}$ represents the directed edge from vertex $v_i$ to vertex $v_j$. The set of all vertices is given by $V=\{v_j\}$ for $j = 1, 2, \dots, N$ for a graph of size $N$ vertices. We define the subgraph $H_j$ as the (inverted) tree graph consisting of all vertices $v_i$ with directed edges into $v_j$. In other words, this represents the subgraph of all incoming directed connections into $v_j$.

We write $H_j(v_{i})$ to denote the set of all vertices $v_i$ in $H_j$, and $H_j[v_{i}]$ to refer to a specific $v_i \in H_j(v_{i})$. Signals traveling on the edges $e_{ij}$ propagate at finite speeds $s_{ij}$, where $0 < s_{ij} < \infty$. As a first approximation, we assume that a single arriving discrete signal has the potential to activate $v_j$. To indicate causality, we use the notation $H_j[v_i] \leadsto v_j$ to signify that vertex $v_i \in H_j(v_i)$ causally activates $v_j$, meaning that $v_i$ is the \emph{winning} vertex whose signal successfully activates $v_j$. We first treat and analyze this restricted (single activation) case and then generalize the model to a latency-dependent spatial-temporal version of a perceptron that activates due to the fractional summation of many arriving discrete signals (section \ref{sec:fracsum}).  

\subsubsection{Refractory period and signal dynamics} \label{sec:prelims}
Each node $v_j$ is assigned an \emph{absolute refractory period}, given by $R_j$, which represents the amount of time the node requires for its internal dynamics to decide its output after being activated. Alternatively, $R_j$ can also reflect a reset period during which $v_j$ cannot respond to subsequent incoming signals. 

We impose no restrictions on the internal processes or models that produce $R_j$. But we assume that $R_j$ can be observed or measured and that $R_j > 0$ (i.e., recovery is never instantaneous, though it may be arbitrarily short). This assumption is reasonable for any physically realizable network.

Consider a vertex $v_i$ with a directed edge $e_{ij}$ connecting to a vertex $v_j$. For $v_i$ to signal $v_j$, a discrete signal must propagate along $e_{ij}$ at a finite speed $s_{ij}$. While $s_{ij}$ could be a constant for all edges, this is not required in the general case. Similarly, while nodes in the network may share the same refractory period (i.e., $R_j = R \; \forall v_j \in V$), the framework can accommodate node-specific values of $R_j$. 

If $v_j$ is activated due to a signal from one of the $v_i$ nodes connected into $v_j$, it becomes refractory for a period $R_j$, during which it cannot respond to additional incoming signals. The temporal nature of $R_j$ implies that as time progresses, $R_j$ gradually decreases and eventually decays to zero, at which point $v_j$ can respond to a new input.

\subsubsection{Internal dynamics of \( v_j\)}  \label{sec:internaldynamics}
Let $\mathcal{T}(v_j)$ represent the instantaneous state of a vertex $v_j$ as a function of some internal node model or activation function at an arbitrary observation or measurement time $T_o$. Informally, the internal state can be interpreted as a binary function, which we define as:

\begin{equation} \label{eq:fj}
\mathcal{T}(V) = 
\begin{cases} 
    1, & \text{if $v_j$ can respond to an input,} \\
    0, & \text{if it is refractory to any input.}
\end{cases}
\end{equation}
Note that we wrote $\mathcal{T}(V)$ in equation \ref{eq:fj} as shorthand to represent the operation of $\mathcal{T}(\cdot)$ on all $v_j \in V$, but will write $\mathcal{T}(v_j)$ to indicate its operation on the specific vertex $v_j$. In the simpler single activation case, we assume that the first arriving discrete signal from any of the $v_i$ vertices connected into $v_j$ can activate it. Once this occurs, $v_j$ becomes refractory for a duration $R_j$, during which $\mathcal{T}(v_j) = 0$. Below, we will progressively build out more neurobiologically realistic conditions (that do not rely on just one signaling event activating $v_j$), and we will formalize what we mean by 'can respond to an input' and 'is refractory to an input'.

\subsubsection{Refraction ratio and signal competition}

To compute (in parallel for all vertices) which vertex $v_i \in H_j(v_i)$ causally activates $v_j$ at discrete times, i.e., $H_j[v_{i}] \leadsto v_j$ in the notation we established above, we track the 'position' of propagating signals along edges, and therefore the amount of latency before they reach target vertices, relative to the refractory states of the vertices. This is achieved using the \emph{refraction ratio}, defined as the ratio between the refractory period $R_j$ and the signaling latency $\tau_{ij}$ associated with a discrete signal from vertex $v_i$ to $v_j$:
\begin{equation} \label{eq:R}
\Delta_{ij} = \frac{R_j}{\tau_{ij}} = \frac{R_j \cdot s_{ij}}{d_{ij}}.
\end{equation}

This ratio provides a basis for computing and determining $H_j[v_{i}] \leadsto v_j$ by analyzing $\Delta_{ij}$ for all $v_i \in H_j(v_i)$ at any given moment.

\subsubsection{Bounds and constraints on $\Delta_{ij}$}

We assume several physical constraints inherent to real-world networks that, as a result, impose bounds on $\Delta_{ij}$
\begin{enumerate}[i]
    \item As $R_j \to 0$, $f_j = 1$ at all times, implying an instantaneous recovery from refractoriness, which is physically unrealizable. Hence, $R_j > 0$.
    \item As $R_j \to \infty$, $f_j = 0$ at all times, resulting in no information flow or signaling. Thus, $0 < R_j < \infty$. 
    \item As $\tau_{ij} \to 0$, $\Delta_{ij}$ becomes undefined, corresponding to $d_{ij} \to 0$ (zero edge length) or $s_{ij} \to \infty$ (instantaneous signaling), which are unattainable conditions.
\end{enumerate}

Thus, $\Delta_{ij}$ is restricted to finite signaling dynamics consistent with physical constraints.

\subsubsection{Effective refractory period and temporal offsets} \label{sec:effectiveR}
If $v_j$ becomes refractory before an arbritrary observation time $T_o$, it could be in this state for a residual time less than the full recovery period $R_j$ at $T_o$. For example, if $v_j$ is halfway recovered from its refractory period at the time the node is observed at $T_o$, then the remaining amount of time before it can be activated again will be $R_j/2$, and not the full period $R_j$. This requires accounting for \lq how much time is left' in both the signaling latencies $\tau_{ij}$ and the refractory periods $R_j$ at $T_o$ to compute activation patterns. We define the residual refractory period at $T_o$ as the \emph{effective refractory period}, denoted $\bar{R}_j$, which is central to our model.

We define the effective refractory period $\bar{R}_j$ as:
\begin{equation} \label{eq:rbar}
\bar{R}_j = R_j - \phi_j, \quad \text{where } 0 \leq \phi_j \leq R_j.
\end{equation}

Here, $\phi_j$ is the temporal offset from $R_j$ at $T_o$:
\begin{enumerate}[i.]
    \item $\phi_j = 0$ implies that $v_j$ becomes refractory exactly at $T_o$.
    \item $\phi_j = R_j$ will occur only when $v_j$ is fully recovered and responsive to any input.
    \item $0 < \phi_j < R_j$ implies that $v_j$ is partially recovered but not yet fully responsive.
\end{enumerate}

Similarly, the signaling latency $\tau_{ij}$ is modified relative to $T_o$ due to temporal offsets in signal propagation. This occurs because recall that by definition $\tau_{ij}$ is the \textit{total} time it takes a signal to travel from node $v_i$ to the target node $v_j$. Any signal that is partially along its way at $T_o$  will take less time, and any signal that initiates after $T_o$ is effectively elongated, i.e., takes longer to arrive at $v_j$. We define the \emph{effective latency} as:
\begin{equation} \label{eq:taubar}
\bar{\tau}_{ij} = \tau_{ij} + \delta_{ij}, \quad \text{where } \delta_{ij} \in \mathbb{R}.
\end{equation}

\begin{enumerate}[i.]
    \item $\delta_{ij} > 0$ encodes a delay in signaling initiation after $T_o$.
    \item $-\tau_{ij} < \delta_{ij} < 0$ will result when a signal partially travels along $e_{ij}$ prior to $T_o$.
    \item $\delta_{ij} = 0$ implies a signal initiation occurs exactly at $T_o$.
\end{enumerate}

\subsubsection{Extended refraction ratio}

We can now extend the refraction ratio to incorporate effective refractory periods and effective latencies as
\begin{equation} \label{eq:Rovert}
\Lambda_{ij} = \frac{\bar{R}_j}{\bar{\tau}_{ij}}.
\end{equation}

Including temporal offsets produces a huge combinatorial solution space for determining activation patterns of $v_j$, although a systematic investigation of this space is beyond the scope of this paper. The global dynamics of the network emerge from the local, statistically independent dynamics of each vertex and its subgraph $H_j(v_i)$. Once $v_j$ is activated, it contributes to the activation dynamics of downstream vertices it connects to, contributing, in turn, to the network's dynamic behavior.

\subsection{Analysis of Activation Conditions}  \label{sec:analysis}

Intuitively, the \emph{activation} vertex $v_i \in H_j(v_i)$ that successfully activates $v_j$, i.e., $H_j[v_{i}] \leadsto v_j$, will correspond to the first signaling event to arrive at $v_j$ immediately after $v_j$ ceases to be refractory. This is equivalent to stating that $H_j[v_{i}] \leadsto v_j$ will be achieved by the $v_i$ with the smallest value of $\bar{\tau}_{ij}$ greater than $\bar{R}_j$. This condition ensures activation and can be expressed as $\bar{\tau}_{ij} \rightarrow \bar{R}_j^+$, i.e., $\bar{\tau}_{ij}$ approaches $\bar{R}_j$ from the right. Determining this condition requires computing the order of arriving signaling events, effectively implementing an algorithm to compute $H_j[v_{i}] \leadsto v_j$ at $T_o$.

To formalize this, we first define a well-ordered set of refraction ratios (in a set-theoretic sense), $\Lambda_{ij}$. We then consider the condition $\bar{\tau}_{ij} \rightarrow \bar{R}_j^+$ and construct a subset whose elements contain only ratios that guarantee signal arrivals after refractoriness has completed:
\begin{equation} \label{eq:BigLSet}
\Lambda^o_{ij} := \{\Lambda_{ij} : i = 1, 2, \dots, N \, | \, \bar{\tau}_{ij} \rightarrow \bar{R}_j^+\}.
\end{equation}

This implies that every $\Lambda_{ij} \in \Lambda^o_{ij}$ satisfies $\Lambda_{ij} < 1$, which follows from the definition of the ratio in equation~\ref{eq:Rovert}, where the refractory period is in the numerator and the latency is in the denominator. 

We impose a structure on $\Lambda^o_{ij}$ by ordering it with the standard $>$ operator, arranging the ratios from largest to smallest. Since $H_j$ is finite, consisting of a finite number of $v_i$ vertices connecting into $v_j$, $\Lambda^o_{ij}$ is also a finite set. Using $\Lambda^o_{ij}$, we compute $H_j[v_{i}] \leadsto v_j$ as follows.

\subsubsection{Theorems for determining activation vertices}

\begin{theorem} \label{theo:theo1}
Assume vertex $v_j$ has an effective refractory period $\bar{R}_j$ at observation time $T_o$. If $\phi_j \neq R_j$, then the condition $H_j[v_{i}] \leadsto v_j$ is given by the refraction ratio $\Lambda_{ij} \in \Lambda^o_{ij}$ that satisfies:
\begin{equation} \label{eq:rwithphi}
v_i = \lceil \max(\Lambda_{ij}) \rceil.
\end{equation}
If, on the other hand, $\phi_j = R_j$, the condition $H_j[v_{i}] \leadsto v_j$ is determined by:
\begin{equation} \label{eq:minlatency}
\min(\bar{\tau}_{ij}) \quad \forall v_i \in H_j(v_i).
\end{equation}
\end{theorem}

Alternatively, Theorem~\ref{theo:theo1} can be expressed equivalently as:

\begin{theorem} \label{theo:theo2}
Assume vertex $v_j$ has an effective refractory period $\bar{R}_j$ at observation time $T_o$. For each $v_i \in H_j(v_i)$ with associated $\bar{\tau}_{ij}$, the condition $H_j[v_{i}] \leadsto v_j$ is satisfied by:
\begin{equation} \label{eq:rwithnophi2b}
\min \left[ (\bar{\tau}_{ij} - \bar{R}_j) > 0 \right].
\end{equation}
\end{theorem}

Algorithmically, equation~\ref{eq:rwithnophi2b} is more efficient to implement, as it requires computing a simple difference rather than a ratio. This efficiency becomes significant when determining all $H_j \in G(V, E)$ in parallel in a network. Proofs for these theorems can be found in \cite{silva2019}.

\subsection{Extensions Beyond the Basic Construction}

In this section, we summarize several natural extensions of the basic single-activation construction. These extensions improve the neurobiological realism and applicability of the model. We first introduce a probabilistic version of the framework, reflecting the stochastic nature of generating action potentials and associated biological processes. We then discuss an approach for handling inhibitory inputs and their impact on network dynamics. Finally, we summarize a version of the framework that incorporates fractional contributions from summating signals. This addition effectively represents a geometric dynamic extension of the classical perceptron model. It is important to note that these extensions are not mutually exclusive. The framework can be implemented to simultaneously accommodate any or all of these extensions, further enriching the model's dynamic complexity and functional capacity.

\subsubsection{Probabilistic extension} \label{sec:prob}
As developed in the preceding sections, the framework is deterministic, meaning that a single vertex $v_i$ is guaranteed to activate $v_j$ when the condition $H_j[v_i] \leadsto v_j$ is satisfied. In other words, the probability of $v_j$ responding to a \emph{winning} signal from $v_i$ is exactly 1. However, biological neural networks are not deterministically precise and are often influenced by random fluctuations in molecular and cellular signaling processes. These fluctuations may arise due to thermal dynamics, sub-diffusion processes, or molecular stochasticity, such as neurotransmitter vesicles crossing the synaptic cleft or the probabilistic binding of neurotransmitters on the postsynaptic membrane.

We define a probability distribution $P_{ij}$ to represent the likelihood that a winning vertex $H_j[v_i] \leadsto v_j$ successfully activates $v_j$. Let $v_j(P_{ij})$ denote the probability of an effect on $v_j$, such as its activation (but see the below), for a given threshold $P_{threshold}$. The activation condition for $v_j$ based on $P_{ij}$ can then be defined as:
\begin{equation} \label{eq:vj1}
v_j(P_{ij}) = 
\begin{cases} 
    1, & \text{if } H_j[v_i] \leadsto v_j \text{ and } P_{ij} \geq P_{threshold}, \\
    0, & \text{if } H_j[v_i] \leadsto v_j \text{ and } P_{ij} < P_{threshold}.
\end{cases}
\end{equation}

It is important to note that equation~\ref{eq:vj1} does not specify the exact nature of $v_j$'s output when $v_j(P_{ij}) = 1$. This could result in a signal being generated if $v_i$ is excitatory or no signal if $v_i$ is inhibitory (see Section~\ref{sec:inh}). The equation solely indicates that $v_j$ \emph{responds} in some manner to $v_i$'s signal appropriate to the behavior of that class of node.

\subsubsection{Inhibitory inputs into $v_j$} \label{sec:inh}

The framework can also accommodate inhibitory inputs in the following manner: If an \emph{activation} vertex $v_i$ satisfies $H_j[v_i] \leadsto v_j$, then:
\begin{enumerate}[i.]
    \item If the input from $v_i$ is excitatory, $v_j$ generates an output signal and becomes refractory for a duration $\bar{R}_j$.
    \item If the input from $v_i$ is inhibitory, $v_j$ does not generate an output signal but becomes refractory for $\bar{R}_j$.
\end{enumerate}

In the inhibitory case, the absence of an output from $v_j$ prevents its contribution to the activation of its downstream vertices while ensuring that $v_j$ remains unable to respond to further inputs until its refractory period ends. 

This approach to modeling excitatory and inhibitory inputs is straightforward and computationally efficient, though other methods for differentiating between input types could be considered.

\subsubsection{Fractional summation of contributing signaling events} \label{sec:fracsum}
In this section, we describe an extension of the framework that incorporates the summation of fractional contributions from multiple nodes towards activating a vertex $v_j$. Instead of a single node $v_i$ solely deterministically or probabilistically activating $v_j$, we consider what happens when we have a \emph{running summation} of contributions from multiple $v_i \in H_j(v_i)$ that collectively need to reach a threshold $\Sigma_T$ to activate $v_j$. Once $v_j$ fires, it becomes refractory, as described above.

Conceptually, this extension represents a geometric refractory model of the classical perceptron, where the signaling latencies and geometric considerations of the network play a critical role in determining activation. These perceptrons can be envisioned as having a geometric morphology that accounts for computed latencies along edges (inputs into $v_j$), similar to biological neurons. The interplay between the latencies of discrete signals, the evolving refractory state of $v_j$, and the decaying contributions of weights determines the dynamic summation towards threshold.

We assume weights and an activation function as in the standard perceptron model. Additionally, we introduce a \emph{decay function} to provide a memory or history for previous signals. This function assigns diminishing but non-zero contributions to the summation from inputs that arrived before the observation time $T_o$. As such, the computational goal is no longer to predict which $v_i \in H_j(v_i)$ will activate $v_j$, but rather which \emph{subset} of $H_j(v_i)$ will collectively do so.

\noindent \textit{Running summation towards threshold.} Assume an observation time $T_o$, and let $v_j$ be non-refractory. The running summation $\Sigma_r$ from $H_j(v_i)$ must reach a threshold $\Sigma_T$ for $v_j$ to activate at some $t \geq T_o$. Upon activation, $v_j$ becomes refractory for a period $R_j$. The specific contribution from each $v_i \in H_j(v_i)$ is given by a weight $w_{ij}$, representing synaptic strength. While not strictly necessary, we assume the added condition that $\Sigma_T$ is constant for all $v_j \in V$.

We assume a set of weights $W_j = \{w_{ij}\}$ for all connections into $v_j$. The maximum weight $w_{ij,max}$ occurs when the signal from $v_i$ first arrives at $v_j$ while it is non-refractory, i.e., at $(\bar{\tau}_{ij} - \bar{R}_j)$ (\emph{c.f.} Theorem~\ref{theo:theo2}). After arrival, we assume a time-varying weight $w_{ij}(t)$ whose contribution decays over time, such that:
\[
w_{ij}(t) < w_{ij,max} \quad \text{for } t > (\bar{\tau}_{ij} - \bar{R}_j).
\]
This finite memory models the progressively diminishing impact of prior signals.

\noindent \textit{Incorporating decay dynamics.} We then need to define a decay function $D_i(t)$ that modulates each weight after its corresponding signal arrives. This ensures that more recent arriving signals contribute more to the running summation than prior signals. At the moment a signal from $v_i$ arrives at $v_j$, the decay function satisfies:
\begin{subequations}
\begin{align}
D_i(\bar{\tau}_{ij} - \bar{R}_j) &= 0, \\
D_i[(\bar{\tau}_{ij} - \bar{R}_j) + \xi] &= 1,
\end{align}
\end{subequations}
where $\xi$ represents the decay duration. For intermediate times:
\begin{equation}
\begin{split}
\text{If } (\bar{\tau}_{ij} - \bar{R}_j) < t < (\bar{\tau}_{ij} - \bar{R}_j) + \xi, \\
\text{then } D_i(t_n) < D_i(t_m) \quad \text{for } t_n < t_m.
\end{split}
\end{equation}

The decay function $D_i(t)$ is strictly increasing, ensuring a progressive reduction in the contribution of $w_{ij}$. The decayed weight is given by:
\begin{equation} \label{eq:w_decay}
w_{ij}(t) = w_{ij,max} - w_{ij,max} \cdot D_i(t),
\end{equation}
where the domain of $D_i(t)$ is $(\bar{\tau}_{ij} - \bar{R}_j) \leq t \leq (\bar{\tau}_{ij} - \bar{R}_j) + \xi$, and its codomain is $0 \leq D_i(t) \leq 1$.

\noindent \textit{Summation with decaying contributions.} 
To compute the fractional contributions of weights towards $\Sigma_r$, consider the subset $\Lambda_M \subset \Lambda^o_{ij}$, where:
\[
\Lambda_M = \{(\Lambda_{ij})_m \in \Lambda^o_{ij} : m = 1, 2, \ldots, M\}.
\]
This subset is the set of $M$ signals (as refraction ratios) that arrive after $v_j$ has recovered from its refractory period that contribute to the running summation that gets $v_j$ to its next activation (firing) state.

The running summation of the maximum weights that contribute to getting $v_j$ to its threshold is:
\begin{equation} \label{eq:w1}
\Sigma_r = \sum_{m=1}^M (w_{ij,max})_m \geq \Sigma_T.
\end{equation}

To account for weight decays as defined above, we compute the fractional contribution of each weight at the moment $t = (\bar{\tau}_{Mj} - \bar{R}_j)$ when threshold is reached:
\begin{equation} \label{eq:w2}
\Sigma_r = \sum_{m=1}^M \left[(w_{ij,max})_m - (w_{ij,max})_m \cdot D_i(\bar{\tau}_{Mj} - \bar{\tau}_{ij})\right] \geq \Sigma_T.
\end{equation}

Here, $D_i(\bar{\tau}_{Mj} - \bar{\tau}_{ij})$ reflects the decay function's value at the time difference between signal arrival and threshold activation. Excitatory and inhibitory inputs can be incorporated as described in Section~\ref{sec:inh}, where inhibitory weights contribute negatively to $\Sigma_r$. Note that in computing equation \ref{eq:w2}, transient computations (e.g., weight decay levels and signaling latencies) do not necessarily need to be stored long-term. This reduces the computational overhead and memory and storage requirements if the evolving dynamics do not need to be measured or visualized until $T_o$.

With the model now fully developed, we can revisit the condition for the state of node $v_j$, given by $\mathcal{T}(V)$, as informally introduced in equation \ref{eq:fj}. The state of $v_j$ at an arbitrary observation time $T_o$ is given by  
\begin{equation} \label{eq:f}
    \mathcal{T}(V) = 
    \begin{cases}
      &\text{if} \; \Sigma_r \geq \Sigma_T, \; v_j = 1\\
      &\text{if} \; \Sigma_r < \Sigma_T, \; v_j = 0\\
    \end{cases}  
\end{equation}
Determining the state given by equation \ref{eq:w2} for every node $v_j$ in the network (with node set $V$) by computing equation \ref{eq:w2} in parallel at an observation or measurement time $T_o$ defines the network dynamical state at $T_o$. 

\vspace{1em}
\noindent\fbox{\parbox{\dimexpr \linewidth-2\fboxsep-2\fboxrule}{\textbf{Box. 1:} Assuming an excitatory network, i.e., a network where for all nodes if a node is activated it generates an excitatory output, we are now in a position to formally restate the key question we ask in this paper: Given some (observed or unobserved) period of evolution dynamics of a network computed by equation \ref{eq:w2} for all nodes $v_j$ in $G(V,E)$, can the dynamics (state) of the network sustain inherent activity beyond an arbitrary observation time $T_o$, or are the dynamics guaranteed to stop, either due to 'epileptic'-like saturation or quiescence? The (global) state of the network is computed by $\mathcal{T}(V) \forall v_j \in V$ (equation \ref{eq:f}), where $\Sigma_r$ is given by equation \ref{eq:w2}.}} 
\vspace{1em}

\section{Preparing Network States for Deutsch-Jozsa Computations}
To leverage the Deutsch–Jozsa algorithm, we need to structure the problem formalized in Box 1. so that it naturally exploits the algorithm's deterministic superposition properties. This will allow us to determine the global state of the network all at once, rather than needing to (classically) compute and check the state of every $v_j$ individually.

To achieve this, we will first construct a specific set that encodes the states of $v_j \in V$. When presented as an input to Deutsch–Jozsa, the output can be interpreted as an answer to the question about the network's dynamics. We will first use Grover's algorithm to evaluate set membership within this framework.

\subsection{Set Membership Conditions} \label{sec:setmem}
Each vertex $v_j \in V$ is associated with a value $\Sigma_r \in \mathbb{R}$ at time $T_o$ that is typically normalized to a defined range:
\begin{equation} \nonumber
    \epsilon \leq \Sigma_r \leq (\epsilon + \Delta),
\end{equation}
commonly within $0 \leq \Sigma_r \leq 1$.

This range can be approximated by an $n$-bit string representation with sufficient precision:
\begin{equation} \nonumber
    \Sigma_r \in \{0,1\}^n.
\end{equation}

For the remainder of the paper, $\Sigma_r$ will refer to this $n$-bit string representation. We do not need to distinguish between the $n$-bit string and continuous numerical representations in the notation. 

In this context, $\mathcal{T}(V)$ is a function that compares the $n$-bit input $\Sigma_r$, the running summation for each vertex $v_j$, to a constant threshold $\Sigma_T \in \{0,1\}^n$ (equation \ref{eq:f}). To determine the global state of the network, we evaluate $\mathcal{T}(V)$ for all $v_j \in V$ at time $T_o$, where the output is either $0$ or $1$ for each node, indicating whether $v_j$ fires (activates) at the next time step (\textit{c.f.} Section~\ref{sec:fracsum}).

For realistic networks, both biological and artificial, the cardinality of $V$, $|V| = N$, may be very large. However, the set of $\Sigma_r$ values is much smaller. Given an $n$-bit precision to represent $\Sigma_r$, many nodes may share the same $\Sigma_r$ values, while some values of $\Sigma_r$ may appear uniquely only once, and others may not occur at all. Thus, only a subset of the possible $n$-bit strings representing the full set of possible values of $\Sigma_r$ may be realized during the network's dynamics.

Mathematically, we construct an ordered list $\mathbb{V}_n$ whose elements map each vertex $v_j \in V$ to its corresponding value of $\Sigma_r$:
\begin{equation} \label{eq:listV}
    \mathbb{V}_n: v_j \rightarrow \Sigma_r \quad \forall v_j \in V.
\end{equation}
Note how this list \emph{may} contain replicate values of $\Sigma_r$.

To formalize this, let $\mathbb{S}_n$ denote the set of all possible $n$-bit strings:
\begin{equation} \label{eq:fullS}
    \mathbb{S}_n \coloneqq \{s_i \,|\, s_i \in \{0,1\}^n\}.
\end{equation}

We then define the following subsets:
\begin{enumerate}[i.]
    \item $S_n$ is the subset of $\mathbb{S}_n$ that represents all $n$-bit string values of $\Sigma_r$ that occur in the network up to the observation time $T_o$, i.e. values of $\Sigma_r \in \mathbb{S}_n$ that are actually realized:
    \begin{equation} \label{eq:Sn}
        S_n \coloneqq \{s_i \,|\, s_i \in \mathbb{S}_n \,\wedge\, s_i = \Sigma_r \in \mathbb{V}_n\}.
    \end{equation}

    \item $S_n^c$ is the complement of $S_n$, containing all $n$-bit strings that do \emph{not} represent values of $\Sigma_r$:
    \begin{equation} \label{eq:Sc}
        S_n^c \coloneqq \{s_i \,|\, s_i \in \mathbb{S}_n \,\wedge\, s_i \neq \Sigma_r \in \mathbb{V}_n\}.
    \end{equation}
\end{enumerate}
Clearly, $S_n \cup S_n^c = \mathbb{S}_n$.

There exists a surjective mapping $\mathbb{V}_n \rightarrow \mathbb{S}_n$: Each $v_j \in V$ maps to one of the $n$-bit string values $s_i \in \mathbb{S}_n$. In general, we expect:
\begin{equation} \nonumber
    |\mathbb{V}_n| = |V| = N > |\mathbb{S}_n| \geq |S_n|.
\end{equation}

For instance, while $|V|$ may be in the millions or billions, representing $\Sigma_r$ to two decimal places requires only 7 bits, encoding 128 values. This includes 101 values from $0.00$ to $0.99$ and the value $1.00$.

Given this construction, we can use Grover's algorithm to determine set membership in $S_n$. The interpretation of $S_n$ in the context of how we framed the neural dynamics question (Box 1.) motivates the construction of the unitary function to be evaluated by the Deutsch–Jozsa algorithm later on and the workflow associated with how we use Deutsch–Jozsa (non-conventionally) so that we can interpret the output of two back to back runs of the algorithm to solve the problem. We progressively develop these ideas over the next few sections.

\subsection{Functional Interpretations of $\mathcal{T}(V)$ Outcomes}

Recall that our goal is to determine if the network's dynamics can sustain inherent activity beyond $T_o$, or whether it is guaranteed to stop due to either 'epileptic' saturation or quiescence. This is achieved by evaluating the global state of the network using $\mathcal{T}(V)$ for each node in the network (equation \ref{eq:f}). From a computational perspective, given how we constructed $S_n$ \emph{from} $\mathbb{V}_n$ (equation \ref{eq:Sn}), the statement in the preceding sentence is equivalent to evaluating $\mathcal{T}(S_n) \forall s_i \in S_n$, where as above for $\mathcal{T}(V)$, we will write $\mathcal{T}(s_i)$ to indicate the evaluation of a specific $s_i$ member of $S_n$ (see the discussion immediately following equation \ref{eq:f}). Formally:
\begin{equation} \label{eq:fSn}
    \mathcal{T}(S_n) = 
    \begin{cases}
      &\text{if} \; s_i \geq \Sigma_T, \; s_i = 1\\
      &\text{if} \; s_i < \Sigma_T, \; s_i = 0\\
    \end{cases}  
\end{equation}

Given how we structured the network dynamics question and how the state of the network is computed at the observation $T_o$ using equation \ref{eq:f} (Box. 1), we can now also (re)interpret the state of the network by appealing to the evaluation of $\mathcal{T}(S_n)$:
\begin{enumerate}[i.]
    \item \emph{Epileptic saturation} is a state where all neurons in the network fire simultaneously at $T_o$, characterized by
    \begin{equation} \nonumber
        \mathcal{T}(S_n) = 1 \quad \forall s_i \in S_n
    \end{equation}

    \item \emph{Quiescence} is a state where no neurons fire, implying no further activity without external input, which occurs when
    \begin{equation} \nonumber
        \mathcal{T}(S_n) = 0 \quad \forall s_i \in S_n
    \end{equation}
In both cases, $\mathcal{T}(S_n)$ is constant (in the language of Deutsch–Jozsa), mapping $S_n$ to binary outputs $\{0,1\}$. 

    \item In contrast, \emph{inherent dynamic activity} will occur only when $\mathcal{T}(S_n)$ is \textit{not} constant, which we can interpret as the network being able to sustain autonomous activity for at least one additional time step. 
\end{enumerate}

The above interpretations allow us to set up a quantum computation whose output resolves the question posed in Box. 1 in the following way:

\vspace{1em} 
\noindent\fbox{\parbox{\dimexpr \linewidth-2\fboxsep-2\fboxrule}{\textbf{Box. 2:} By running the Deutsch–Jozsa algorithm twice, we can distinguish between constant and non-constant states of $\mathcal{T}(S_n)$. A non-constant result implies that the network is neither epileptic nor quiescent. It is important to note that $\mathcal{T}(S_n)$ need not be balanced — only not constant. This agrees with the Deutsch–Jozsa requirement for unitary functions of the form $f(\{0,1\}^n) \rightarrow \{0,1\}$. \newline

\textbf{It is \textit{critical} to note that this is a special use case of the Deutsch-Jozsa algorithm.} Deutsch-Jozsa takes as input a function that evaluates a binary sequence of 0's and 1's, i.e., a set $\mathbb{S}_n = \{0,1\}^n$, and uses quantum superposition to simultaneously determine if the function's output evaluated on the $n$-bit set is \textit{constant} (all 0's or all 1's) or whether the output is \textit{balanced} (an equal number of 0's and 1's). However, in our case here, we need to evaluate $\mathcal{T}(S_n)$ (equation \ref{eq:fSn}) \emph{only} on the subset $S_n \subset \mathbb{S}_n$ using Deutsch-Jozsa (\emph{c.f.} equation \ref{eq:Sn}). But the algorithm cannot evaluate a particular subset of $\{0,1\}^n$. Thus, it cannot explicitly evaluate $\mathcal{T}(S_n)$ directly. Thus, we first need to extend $S_n$ to a construction compatible with the Deutsch-Jozsa, which assumes that the oracle function being evaluated is defined over the entire domain $\mathbb{S}_n$.  We show how to carry this out below.
}}
\vspace{1em} 

\subsection{Determining Membership in $S_n$ and $S_n^c$ Using Grover's Algorithm} 
To identify which elements of $\mathbb{S}_n$ belong to $S_n$ given the dynamics encoded in $\mathbb{V}_n$, we need to search for each $n$-bit string $s_i \in \mathbb{S}_n$ that is in the list $\mathbb{V}_n$ which contains the values of $\Sigma_r$ resulting from the dynamic evolution of the network up to the observation time $T_o$. The resultant set is $S_n$, and its complement in $\mathbb{S}_n$ forms $S_n^c$. Recall that we expect some elements of $\mathbb{S}_n$ to appear in $\mathbb{V}_n$ only once, some multiple times, and others not at all. Grover's algorithm offers a potential quadratic speedup over classical approaches to construct $S_n$ by enabling an efficient search within $\mathbb{S}_n$. In particular, for our purposes here, we are effectively using Grover to search for the values in $\mathbb{S}_n$ that belong to $S_n$ in order to \lq mark' them with a phase-flip in a quantum circuit, and not necessarily to amplify them. This will be necessary because the members of $S_n$ will need to be functionally evaluated by equation \ref{eq:fSn} using a bit-subtraction approach outlined below. The elements of the complement $S_n^c$ will not be evaluated by equation \ref{eq:fSn} and will instead be assigned a constant value of $0$ or $1$ for each of the two Deutsch-Josza runs. The practical reasons for leveraging Grover for this task become evident when attempting to implement this model in a series of quantum circuits. We return to this below in Section \ref{ssec:grover-multi-marked}.

\subsubsection{Construction of the oracle function}

The list $\mathbb{V}_n$ contains the running summation values $\Sigma_r$ for each vertex $v_j$ at $T_o$ (see equation~\ref{eq:listV}). Assume each element in $\mathbb{V}_n$ is labeled as $\{\nu_1, \nu_2, \dots, \nu_M\}$. Each of these elements of $\mathbb{V}_n$ corresponds to an $n$-bit string in $\mathbb{S}_n$. For each $s_i \in \mathbb{S}_n$, we construct an oracle function $U_f$ to identify if $s_i \in \mathbb{V}_n$. Specifically, $U_f$ checks if there exists a $\nu_i \in \mathbb{V}_n$ such that $\nu_i = s_i$ for some $s_i \in \mathbb{S}_n$:
\begin{equation}
    \exists \nu_i \in \mathbb{V}_n \text{ such that } \nu_i = s_i.
\end{equation}

The oracle function $U_f$ performs the operation:
\begin{equation}
    U_f\ket{\nu_i} = (-1)^{f(\nu_i)}\ket{\nu_i},
\end{equation}
where $\ket{\nu_i}$ encodes a state corresponding to $\nu_i \in \mathbb{V}_n$, and $f(\nu_i)$ evaluates to 1 if $\nu_i = s_i$, and 0 otherwise.

To construct $U_f$, we will use a standard gate-based construction of Grover that implements the logical condition \lq if the current state $\nu_i = s_i$, apply a phase flip; otherwise, do nothing\rq (Section \ref{ssec:grover-multi-marked}).

\subsubsection{Diffusion operator and search space}

The standard diffusion operator $D$ in Grover's algorithm is defined as:
\begin{equation}
    D = 2\ket{\sigma}\bra{\sigma} - I,
\end{equation}
where $\ket{\sigma}$ is the uniform superposition over all $s_i \in \mathbb{S}_n$:
\begin{equation} \label{eq:searchspace}
    \ket{\sigma} = \frac{1}{\sqrt{N}} \sum_{s_i=0}^{N-1} \ket{s_i}.
\end{equation}
Here, $I$ is the $N \times N$ identity matrix, and $N = |\mathbb{S}_n| = 2^n$. The diffusion operator amplifies the amplitudes of marked states identified by $U_f$, inverting their amplitudes about the average. In our context, Grover's algorithm iteratively applies $U_f$ and $D$ to mark $s_i \in \mathbb{S}_n$ that are present in $\mathbb{V}_n$, which will allow efficient identification of elements in $S_n$ (see the discussion below in Section \ref{ssec:grover-multi-marked}).

\subsubsection{Iterative application of Grover's algorithm}

To determine membership of each $s_i \in \mathbb{S}_n$, the algorithm has to be run iteratively for each $s_i$. For example, continuing the example from above, a practically reasonable precision of $\Sigma_r$ requiring at most 128 $n$-bit strings (7-bit precision) would involve $|\mathbb{S}_n| = 2^n \approx 10^2$ iterations.

In general, the number of iterations needed for amplitude amplification for a specific $s_i$ is approximately is
\begin{equation} \nonumber
    k \approx \frac{\pi}{4}\sqrt{\frac{N}{\mu}},
\end{equation}
where $\mu$ is the number of occurrences of $s_i$ in $\mathbb{V}_n$. The algorithm's efficiency improves with higher $\mu$ values because fewer iterations are required. For real physically constructable networks, there are likely other physical and functional constraints. So optimizing $k$ may benefit from structural insights into $G(V,E)$, which could suggest and constrain patterns in $\mathbb{V}_n$. For example, signaling conditions that constrain the possible range of values $\Sigma_r$ can take. In practicality, though, while we can successfully mark the states $s_i \in S_n$, amplification is more complex than this because we are applying Grover to multiple states simultaneously. Again, see the detailed discussion in in Section \ref{ssec:grover-multi-marked} below in the paper. 

In the worst case, the algorithm requires approximately $\frac{\pi}{4}\sqrt{N}$ iterations, yielding a quadratic speedup over classical search methods, which would require $O(N)$ steps to evaluate each $s_i$ sequentially.

Once $S_n$ is identified and marked, determining $S_n^c$ is straightforward. It is simply the complement of $S_n$ in $\mathbb{S}_n$: $S_n^c = \mathbb{S}_n \setminus S_n$. This allows us to construct both subsets efficiently using Grover, setting up the computations we will do in the Deutsch–Jozsa computations.

\section{Evaluating Network Dynamics with the Deutsch-Jozsa Algorithm} \label{sec:DJ}
With $S_n$ and $S_n^c$ defined, we can now use the Deutsch–Jozsa algorithm to determine if the network can sustain inherent recurrent activity beyond $T_o$, or if its dynamics will necessarily cease. However, as emphasized in Box 2., the evaluation must be done on $S_n$, a subset of $\mathbb{S}_n = \{0,1\}^n$, rather than on the full domain.

To address this subset evaluation requirement, we extend $S_n$ into $\mathbb{S}_n$ by taking advantage of the relationship $S_n \cup S_n^c = \mathbb{S}_n$. This construction will let us interpret the Deutsch–Jozsa algorithm's output to indirectly infer the behavior of $\mathcal{T}(S_n)$. Specifically, the elements of $S_n$ encode the key information about the network dynamics. 

For elements in $S_n^c$, there is no explicit evaluation. Instead, all elements of $S_n^c$ are assigned a fixed value (either 0 or 1) for one algorithm run and then reassigned to the other fixed value for a second subsequent run. This assignment ensures that we can infer the behavior of $\mathcal{T}(S_n)$ based on the algorithm's output. To do this, we need to construct and evaluate a two-part unitary function $U_f$ across two distinct runs of the algorithm.

\subsection{Evaluating $S_n$ and $S_n^c$ with a Two-Part $U_f$}
We define a two-part oracle function $U_f$ that operates on the composite state $\ket{s_i}\ket{-}$, where the ancillary qubit $\ket{-}$ is initialized as
\begin{equation}
    \ket{-} = \frac{1}{\sqrt{2}}(\ket{0} - \ket{1}).
\end{equation}

\subsubsection{Behavior for $S_n$}
For $s_i \in S_n$, the oracle function evaluates \( \mathcal{T}(s_i) \), flipping the phase of \( \ket{s_i} \) when \( \mathcal{T}(s_i) = 1 \) and leaving it unchanged when \( \mathcal{T}(s_i) = 0 \), such that
\begin{subequations} \label{eq:fsi}
    \begin{align}
        &\text{If } \mathcal{T}(s_i) = 1: \quad U_f\ket{s_i}\ket{-} = -\ket{s_i}\ket{-}, \\
        &\text{If } \mathcal{T}(s_i) = 0: \quad U_f\ket{s_i}\ket{-} = \ket{s_i}\ket{-}.
    \end{align}
\end{subequations}
Note that for $S_n$, this computation will be the same for both runs of the algorithm. 

\subsubsection{Behavior for $S_n^c$}
For $s_i \in S_n^c$, the oracle does not evaluate \( \mathcal{T}(s_i) \) directly. Instead:
\begin{enumerate}[i.]
    \item For the first run of the algorithm, all elements \( s_i \in S_n^c \) are assigned a value of 0, i.e., $\mathcal{T}(S_n^c)= 0$ always.
    \item In a separate second run, all elements \( s_i \in S_n^c \) are assigned a value of 1, i.e., $\mathcal{T}(S_n^c)= 1$ always..
\end{enumerate}
This ensures that \( S_n^c \) does not contribute to the dynamic information encoded in \( \mathcal{T}(S_n) \). But it satisfies the technical requirement that any input function $f(\cdot)$ into Deutsch-Jozsa operates over the entire domain $\mathbb{S}_n$ by filling in the values of $\mathbb{S}_n$ not encountered in $S_n$ that encodes the dynamics.

Mathematically, for \( s_i \in S_n^c \):
\begin{subequations} \label{eq:fsic}
    \begin{align}
        &\text{If assigned } \mathcal{T}(s_i) = 1: \quad U_f\ket{s_i}\ket{-} = -\ket{s_i}\ket{-}, \\
        &\text{If assigned } \mathcal{T}(s_i) = 0: \quad U_f\ket{s_i}\ket{-} = \ket{s_i}\ket{-}.
    \end{align}
\end{subequations}

\subsection{Constructing $U_f$ to Evaluate $f(S_n)$} \label{sec:binsum}
To construct $U_f$, we need to combine two operations:
\begin{enumerate}[i.]
    \item Determination of set membership: Mark the elements of $S_n^c$ so they are assigned fixed values without explicit evaluation (equation \ref{eq:fsic}).
    \item Evaluation of $S_n$: Apply \( \mathcal{T}(s_i) \) to elements in $S_n$ according to equation~\ref{eq:fsi}.
\end{enumerate}

The first step requires marking $s_i \in S_n^c$. This can be done using controlled phase-flip operations that selectively identify and act on states corresponding to $S_n^c$, leaving elements of $S_n$ unchanged. These marking operations are performed in parallel across the superposition of $\mathbb{S}_n$. 

The second step evaluates $\mathcal{T}(S_n)$ by comparing each $\Sigma_r$ (encoded in $s_i$) with the threshold $\Sigma_T$. We implement this as a binary subtraction $\Sigma_r - \Sigma_T$, with the ancillary qubit encoding the result
\begin{enumerate}[i.]
    \item \( \ket{1} \) if \( s_i \in S_n = \Sigma_r \geq \Sigma_T \), corresponding to \( \mathcal{T}(s_i) = 1 \).
    \item \( \ket{0} \) if \( s_i \in S_n = \Sigma_r < \Sigma_T \), corresponding to \( \mathcal{T}(s_i) = 0 \).
\end{enumerate}

\subsubsection{Example of a Binary Subtraction for Evaluating $f(S_n)$} \label{sec:bin_check}
Let's work through a specific example of a binary subtraction operation. Consider a vertex $v_j \in G(V,E)$ with a corresponding entry $\nu_i \in \mathbb{V}_n$ that encodes the running summation value $\Sigma_r$ at the observation time $T_o$ for the specific vertex $v_j$. This value would be marked as an element $s_i \in S_n$. Assume a 7-bit representation of $\Sigma_r$ and the threshold $\Sigma_T$ to two-decimal precision, as discussed in Section~\ref{sec:setmem}. 

Let
\begin{equation} \nonumber
    \ket{\nu} = \ket{1010101}, \quad \ket{\tau} = \ket{0110010}.
\end{equation}
where $\ket{\nu}$ encodes $\Sigma_r = 0.85$ in binary, while $\ket{\tau}$ encodes $\Sigma_T = 0.50$. The ancillary qubit, initially in state $\ket{0}$, will store the result of comparing $\Sigma_r \geq \Sigma_T$.

The binary subtraction operation is as follows:
\begin{center}
\begin{tabular}{c c c c c c c c}
$\nu =$ & 1 & 0 & 1 & 0 & 1 & 0 & 1 \\
$\tau = $ & 0 & 1 & 1 & 0 & 0 & 1 & 0 \\
\hline
Result = & 1 & -1 & 0 & 0 & 1 & -1 & 1
\end{tabular}
\end{center}

Starting from the most significant bit, we compare $\rho_6$ to $\tau_6$. In this example, since $\rho_6 = 1 > \tau_6 = 0$, we can immediately determine that $\Sigma_r > \Sigma_T$ without needing to compare any other bits. If necessary, the comparison would progress to the next significant qubit until a determination is made. The consequence here is that the ancillary qubit is flipped to $\ket{1}$ to indicate that $f(s_i) = 1$. 

This binary subtraction comparison is efficient because it focuses only on detecting the sign of the result (i.e., positive or negative) rather than computing the full numerical value of $\Sigma_r - \Sigma_T$. We discuss the thresholding sub-circuit that we used in our numerical simulations and experiments to implement this scheme in section \ref{sec:threshold}.

\subsection{Measurement Interpretations of $U_f$}
By running the Deutsch–Jozsa algorithm twice, assigning all elements of \( S_n^c \) first to \( f(s_i) = 0 \) and then to \( f(s_i) = 1 \), measurement outcomes provide an interpretation that answers our question about the network dynamics (Box 1.):
\begin{enumerate}[i.]
    \item \textbf{Quiescence:} If $S_n^c = 0$ and $U_f$ evaluates to a constant state $\ket{1}^{\otimes n}$, the network is quiescent, because this can only occur when $\mathcal{T}(S_n) = 0$ for all $s_i \in S_n$.
    \item \textbf{Epileptic State:} If $S_n^c = 1$ and $U_f$ evaluates to a constant state $\ket{0}^{\otimes n}$, the network is epileptic, because this can only occur $\mathcal{T}(S_n) = 1$ for all $s_i \in S_n$.
    \item \textbf{Dynamic Activity:} If $U_f$ is not constant for both cases, we interpret this result as the network being able to potentially maintain inherent dynamic activity due to the measurement outcome reflecting what must be a mix of active and inactive vertices in $S_n$.
\end{enumerate}

\section{Comparative Algorithmic Complexity} \label{sec:complexity}
In this section, we analyze the algorithmic complexity of solving the problem using our workflow leveraging the Grover and Deutsch–Jozsa algorithms versus a purely classical approach. We will show that using these quantum algorithms theoretically conveys a significant computational advantage and speedup and demonstrate how such an advantage scales, i.e., becomes greater, as the size of the network being analyzed increases. 

\subsection{Classical Complexity}
Consider a network with $N = |V|$ vertices and let \( S_n \subseteq \mathbb{S}_n \) represent the set of \( n \)-bit encodings of \( \Sigma_r \), where \( |\mathbb{S}_n| = 2^n \). A classical approach would need to

\begin{enumerate}[i.]
    \item Evaluate \( \mathcal{V} \) for each \( v_j \in V \) in order to compare \( \Sigma_r \) and \( \Sigma_T \) (equation~\ref{eq:f}). This requires iterating over all vertices. 
    \item Combine the results of for all \( v_j \in V \) to determine whether the network is quiescent, epileptic, or dynamically active.
\end{enumerate}

Assuming a binary comparison operation of complexity \( O(n) \) for each \( v_j \), the total complexity of the classical approach is
\[
O_{\text{classical}} = O(N \cdot n).
\]

\subsection{Quantum Complexity: Grover’s Algorithm}
Grover’s algorithm provides a quadratic speedup for searching an unstructured set. To determine \( S_n \), we use Grover’s algorithm to find all \( s_i \in \mathbb{S}_n \) that belong to \( S_n \). Let \( \mu \) denote the number of occurrences of a particular \( s_i \) corresponding to a a value \( \Sigma_r \in \mathbb{V}_n \), and \( |\mathbb{S}_n| = 2^n \). The number of iterations required for Grover’s algorithm to amplify the amplitude of each solution is approximately
\[
k \approx \frac{\pi}{4} \sqrt{\frac{2^n}{\mu}}.
\]

Assuming the worst case where \( \mu = 1 \), the complexity of finding each \( s_i \in S_n \) is given by
\[
O_{\text{Grover}} = O\left(2^{n/2}\right).
\]

If \( S_n \) contains \( |S_n| \) elements, the total complexity for constructing \( S_n \) is
\[
O_{\text{Grover (total)}} = O\left(|S_n| \cdot 2^{n/2}\right).
\]

\subsection{Quantum Complexity: Deutsch–Jozsa Algorithm}
The Deutsch–Jozsa algorithm evaluates whether \( \mathcal{T}(S_n) \) is constant or not across all elements of \( \mathbb{S}_n \). Unlike a classical approach that would require checking each \( s_i \in \mathbb{S}_n \), the Deutsch–Jozsa algorithm achieves this in a constant number of evaluations:
\[
O_{\text{Deutsch-Jozsa}} = O(1).
\]

When combined with Grover’s algorithm for constructing \( S_n \), the total quantum complexity becomes
\[
O_{\text{quantum total}} = O\left(|S_n| \cdot 2^{n/2} + 1\right).
\]

\subsection{Comparative Analysis}

As such, the task of evaluating \( \mathcal{T}(S_n) \) for all vertices in \( V \) can be summarized by
\begin{enumerate}[i.]
    \item \textit{Classical complexity:}
    \[
    O_{\text{classical total}} = O(N \cdot n).
    \]
    \item \textit{Quantum complexity:}
    \[
    O_{\text{quantum total}} = O\left(|S_n| \cdot 2^{n/2} + 1\right).
    \]
\end{enumerate}

For large \( N \) and \( n \), a quantum algorithmic approach provides a significant increase in computational efficiency. Grover’s algorithm provides a quadratic speedup in \( 2^n \), while Deutsch–Jozsa evaluates \( \mathcal{T}(S_n) \) in constant time. 

Let's consider a numerical example to illustrate this advantage. Consider a network with \( N = 10^6 \) vertices, where, as we have done throughout the paper, \( n = 7 \) bits are used to represent \( \Sigma_r \) with precision sufficient for two significant digits. This means that \(|\mathbb{S}_n| = 2^n = 128 \). Assume that \( |S_n| = 100 \) .

For the \textbf{classical approach} each vertex requires \( O(n) = O(7) \) steps for comparison. Evaluating all \( N \) vertices gives
\[
O_{\text{classical total}} = O(10^6 \cdot 7) = O(7 \times 10^6).
\]

For the \textbf{quantum approach}, using Grover to find all elements of \( S_n \), with \( |S_n| = 100 \) and \( |\mathbb{S}_n| = 128 \) will require approximately
\[
k \approx \frac{\pi}{4} \sqrt{\frac{128}{1}} \approx 11.3.
\]
number of iterations. Thus, the total cost for Grover’s step will be
\(
O_{\text{Grover (total)}} = O(100 \cdot 11.3) \approx O(1130).
\)

Subsequently evaluating \( \mathcal{T}(S_n) \) with Deutsch–Jozsa adds a constant overhead of
\(
O_{\text{Deutsch-Jozsa}} = O(1).
\)

As a result, the total quantum complexity is given by
\(
O_{\text{quantum total}} \approx O(1130 + 1) = O(1131).
\)

\paragraph{Speedup:}
We can assess the comparative theoretical speedup advantage of the quantum approach versus the classical approach by computing the ratios of their respective complexities. The classical approach needs \( O(7 \times 10^6) \), while the quantum approach requires \( O(1131) \). This then translates to a speedup factor of approximately
\[
 \frac{7 \times 10^6}{1131} \approx 6188.
\]
This means that the quantum approach is over 6000 times more computationally efficient than the classical approach for this problem. 

By \lq faster\rq, we specifically mean that a quantum approach, the way we describe it here, requires significantly fewer computational steps to arrive at the result. This reduction translates into less computational time, assuming similar hardware clock speeds and negligible algorithmic overhead. This also assumes a physical quantum computer with sufficiently stable logical qubits. 

In general, as the size of the network, \( N = |V| \), increases, the computational advantage provided by the quantum approach becomes increasingly significant. 

We can approximate a comparative scaling that shows how the speedup magnifies as \( N \) increases:
\[
\sim \frac{O(N \cdot n)}{O(|S_n| \cdot 2^{n/2} + 1)}.
\]
For large \( N \), the classical complexity \( O(N \cdot n) \) grows linearly, while the quantum complexity remains tied primarily to \( |S_n| \) and \( 2^{n/2} \). This growing gap results in an increasingly significant computational advantage for the quantum approach.

For large-scale networks typical in the brain or artificial neural networks, \( N \) can be many billions of vertices. The classical approach will eventually become computationally infeasible due to its linear dependence on \( N \), while the quantum approach remains algorithmically efficient. This scalability, at least in theory, emphasizes the potential of a quantum approach like the one we developed here for analyzing the dynamics of large, high-dimensional networks.


\section{Quantum Circuit Implementation and Numerical Simulations}
In order to make the theoretical ideas more concrete, we constructed a straightforward series of quantum circuits and ran numerical simulation toy examples. Our test code was written in Python using Google's Cirq quantum circuit design library and executed in Google's qsim quantum simulator (\url{https://quantumai.google/qsim}) via a Colab notebook (\url{https://colab.research.google.com/)}. The code and all supporting documentation are available on GitHub: \url{https://github.com/gabe-alex-silva/Network\_Dynamics\_QuantumSim}.

\subsection{Quantum Circuit and Workflow Overview and Simulation Set Up} \label{sec:simsetup}
We assumed a (classical) network consisting of six nodes with evolving dynamics determined by the fractional summation model outlined in section \ref{sec:fracsum} up to some arbitrary observation time $T_o$. At $T_o$, each of the six nodes $v_j$ was assigned a running summation $\Sigma_r$ represented as a 7-bit value (\textit{c.f.} section \ref{sec:setmem} above). As such, there would be at most $2^7 = 128$ possible $\Sigma_r$. In other words, $\mathbb{V}_n$, representing the collection of $\Sigma_r$ realized among the six nodes, was taken as arbitrarily chosen subsets of the full range of possible $\Sigma_r$ with 7-bit precision, i.e., $\{0,1\}^7$, to identify and construct $S_n \subset \mathbb{S}_n$. 

The complement of $S_n$, $S_n^c$, was then the subset of \( \{0,1\}^7 \) that was \textit{not} realized by the network dynamics (\textit{c.f.} section \ref{sec:setmem}). We first constructed and simulated a \textbf{Grover sub-circuit} to simultaneously search and mark all 7-bit values in \( \{0,1\}^7 \) that were realized values of \( \Sigma_r \). 

A second \textbf{thresholding sub-circuit then evaluated \( \mathcal{T}(s_i) \) to construct the two-part \( U_f \)}. For each \( s_i \in S_n \) the circuit compared each corresponding value \( \Sigma_r \) to a  threshold \( \Sigma_T \) via the binary subtraction approach we described in section \ref{sec:binsum}, i.e., checking if \( \Sigma_r \geq \Sigma_T \) bit by bit. If \( \Sigma_r \geq \Sigma_T \) the operation marked \( \mathcal{T}(s_i) = 1 \), otherwise, if \( \Sigma_r < \Sigma_T \) it marked \( \mathcal{T}(s_i) = 0 \). For \( s_i \in S_n^c \) no explicit evaluation was carried out. Instead, a constant value of \( \mathcal{T}(s_i) = 0 \) was assigned to the first Deutsch-Josza run, and a constant value of \( \mathcal{T}(s_i) = 1 \) was assigned to the second run. 

Finally, we passed \( U_f \) to a \textbf{ Deutsch-Josza sub-circuit} that ran twice, once for each of the two constant values of \( \mathcal{T}(s_i) \) for all \( s_i \in S_n^c \). $S_n$ was kept the same for both runs, as required, and the outputs of both runs were then compared to arrive at an interpretation of the network dynamics.  

\begin{figure}
  \includegraphics[width=\linewidth]{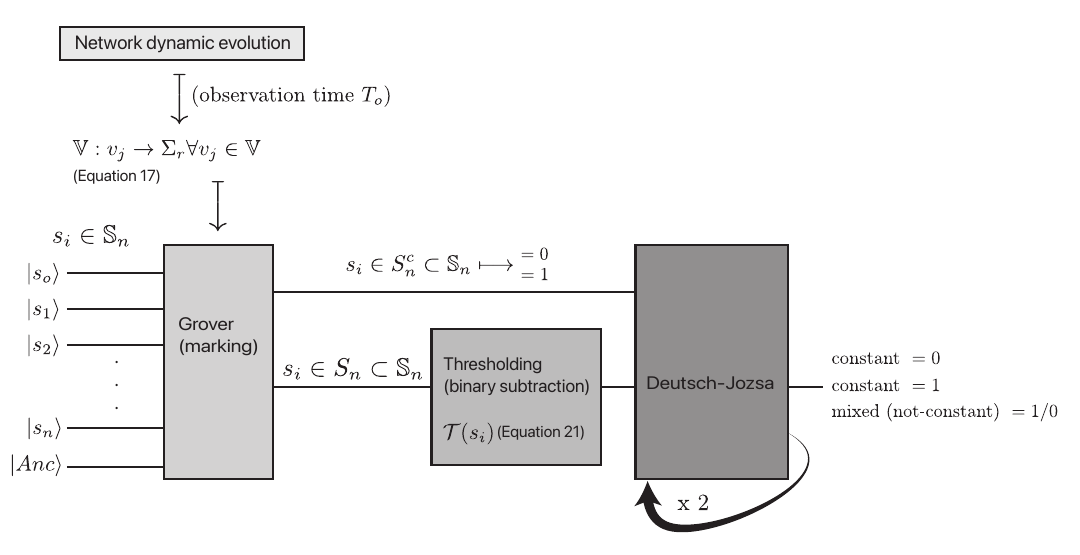}
  \caption{\textbf{Schematic overview of the three main sub-circuits and their workflow.} The \textbf{Grover sub-circuit} places the full input range $\mathbb{S}_n$ into a uniform superposition, applies the marking oracle (phase-flip) on states $s_i \in S_n$ given the state of the network dynamics at the observation time $T_o$ encoded by $\mathbb{V}_n$, and performs a standard diffusion step. One or more iterations of oracle + diffusion can be performed on the marked states. The \textbf{Thresholding sub-circuit} then compares the value of $\Sigma_r$ for each system state $s_i \in \mathbb{S}_n$ to a threshold value $\Sigma_{T}$, via a set multi-controlled $\mathbb{Z}$-gates. If $\Sigma_r \ge \Sigma_{T}$, an ancilla qubit is flipped to encode $\mathcal{T}(s_i)=1$; otherwise $\mathcal{T}(s_i)=0$. Finally, in the \textbf{Deutsch--Jozsa sub-circuit}, all complement states $s_i \in S_n^c$ (non-actualized dynamics) are forced to either 0 or 1, while the threshold logic output is preserved for $s_i \in S_n$. A final Hadamard on the system qubits plus measurement indicates whether the resulting function is effectively constant 0,  constant 1, or a mixture, distinguishing epileptic saturation, quiescence, or continued activity, respectively. (See text for details.) Note that the internal control gate circuit structures for each sub-circuit are not shown in this diagram.}
  \label{fig:fullcirc}
\end{figure}

Figure \ref{fig:fullcirc} shows an integrative schematic of the relationship and chaining of each of the three principle sub-circuits in our workflow. In the rest of this section, we discuss each circuit in detail and end by showing how we can solve the network dynamics problem as theoretically structured in the preceding parts of this paper, illustrating three representative end-to-end simulation experiments and their results for our six node test network. 
 
\subsection{Amplitude Amplification of Multiple Marked States with Grover} 
\label{ssec:grover-multi-marked}
We constructed a standard Grover circuit that put all \( |\mathbb{S}_n| = \{0,1\}^7 =128 \) states in an equal superposition via a set of Hadamard gates operating on each of the 7 qubits. During the oracle's marking step, \emph{only} those states in \( S_n \) are marked with a minus-phase. The subsequent diffusion step inverts amplitudes about the average, causing amplitude to flow into the negatively marked members of \( S_n \). Once the states are marked, iterations of the diffusion cycle increase the amplitude of the marked states at the expense of the unmarked states in \( S_n^c \). So even though all 128 states in \( \mathbb{S}_n \) are held in superposition, only the marked subset was selectively marked and (theoretically) amplified with each iteration. The circuit's construction was standard, but the way we used it in our workflow was not. 

\subsubsection{Leveraging Grover's algorithm} \label{sec:groveradvantage}
There are two practical reasons we leverage Grover in our workflow. First, to implement the thresholding operation and complement logic in our overall process, discussed in detail in the next two sections, it is essential to first identify and mark the states in $S_n$ representing the actualized subset of $\mathbb{S}_n = \{0,1\}^n$ ($n=7$ in our simulations here) encoded by the values of $\Sigma_r \in \mathbb{V}_n$ that represent the network dynamics. We use Grover’s algorithm to perform this search efficiently, marking the relevant states without requiring an exhaustive (classical) enumeration process. The marking step is critical to ensure that the subsequent quantum operations, specifically the thresholding evaluation and Deutsch–Jozsa sub-circuits, are applied correctly. Without Grover, one would need to iterate over all $2^n$ states to identify $S_n$ \lq manually', defeating the purpose of any quantum speedup. By marking $S_n$ in quantum superposition using Grover, we ensure that the thresholding and complement evaluations are applied only to the relevant states, optimizing the approach for a quantum implementation.

The second purpose is more nuanced but sets the theoretical stage for thinking about this class of network dynamics problem beyond Grover and Deutsch-Jozsa. In a \emph{fully} quantum or \lq oracle-based' construction, one can imagine the network’s internal dynamics evolving independent of an outside observer, thereby generating an unknown or hidden set of dynamics encoded by $\mathbb{V}_n: v_j \rightarrow \Sigma_r \quad \forall v_j \in V$ (equation \ref{eq:listV}). This could be due to a classical dynamic evolution model like the one we introduce in this work \emph{or} perhaps \textbf{a novel analog-digital neural dynamic model that more closely resembles how we know biological neurons compartmentalize processes and compute. Biological neurons are \emph{not} purely digital but rather highly complex analog-digital computing elements individually that collectively contribute to network dynamics. An approach that leverages the inherent properties of the Bloch sphere or an evolving Hamiltonian before collapsing to a binary output is a unique way to think about an analog-digital hybrid neural dynamic model.} Such a model could conceivably have applicability to understanding the biological brain and new forms of quantum machine learning and artificial neural networks. In such a model, the evolution of the network may not be observable, and $\mathbb{V}_n$ may need to pass directly to Grover (or some other quantum algorithm) as part of a larger quantum process. In this scenario, Grover’s algorithm would query the dynamics without being classically observed. In practice, the network’s evolution (or an encoding of it) as a black-box oracle would be marked without observation before a final measurement that reveals the actualized network dynamic states with high probability.

\subsubsection{Grover simulation results: Identifying and marking mutiple network states in $S_n$}
\begin{figure}
  \includegraphics[width=\linewidth]{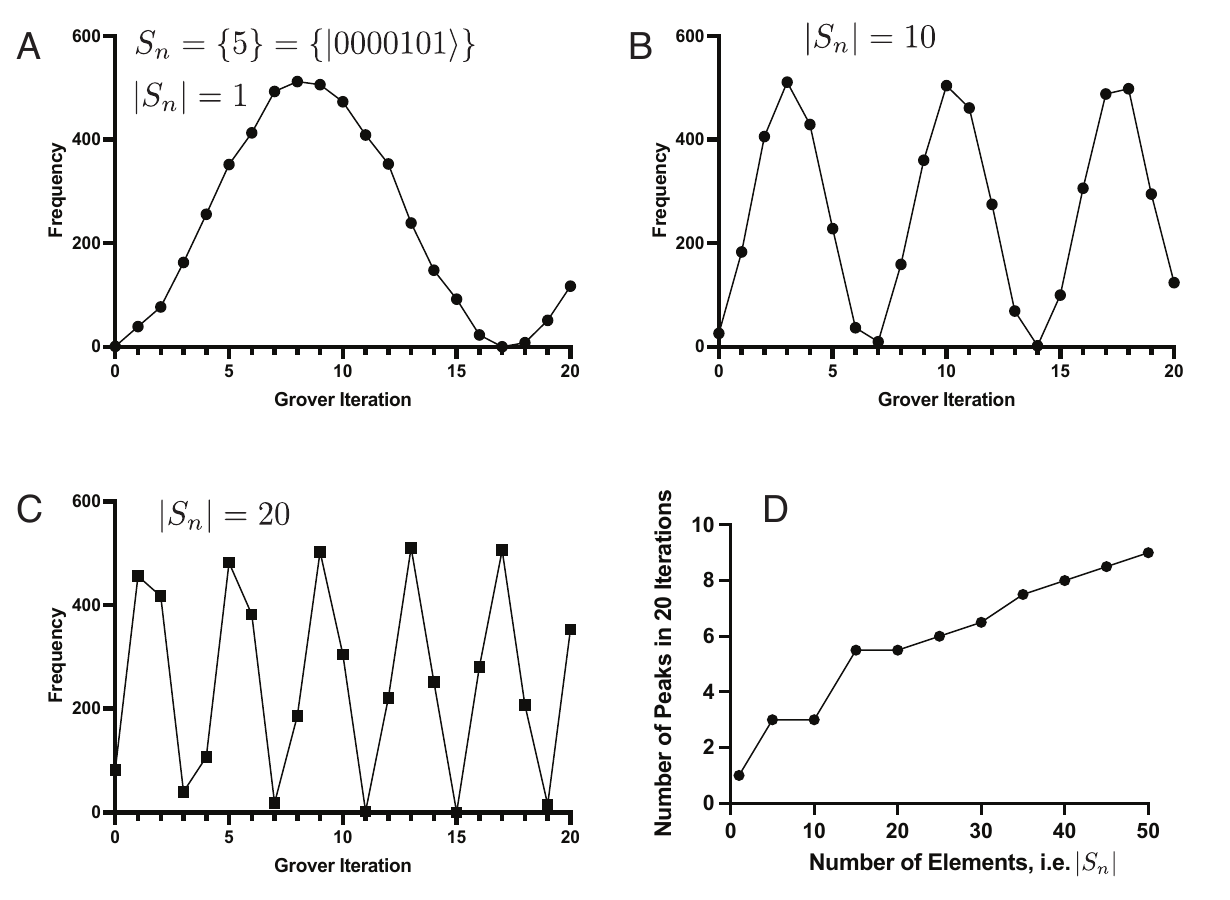}
  \caption{\textbf{Grover algorithm frequency histograms for three representative examples of different sized \(S_n\)}: \(|S_n| = 1\) (\textbf{Panel A}), \(=10\) (\textbf{Panel B}), and \(=20\) (\textbf{Panel C}). All runs consisted of 512 shots over a range between \(0\) and \(20\) Grover iterations, i.e., phase flip marking of network states in \(S_n\) followed the amplitude amplification diffusion step. \textbf{D.} Plot of \(|S_n|\) versus the number of peaks in the frequency histogram over the range range of \(|S_n| \leq 50\). Note how as the number of states in \(S_n\), i.e., the size of \(|S_n|\), increases, the frequency of oscillations of the histogram increases monotonically, as expected.}
  \label{fig:grover}
\end{figure}

Figure \ref{fig:grover} shows the output frequency histogram for three representative examples of different sized \(S_n\): \(|S_n| = 1\) (Figure \ref{fig:grover}A), \(=10\) (Figure \ref{fig:grover}B), and \(=20\) (Figure \ref{fig:grover}C). All runs consisted of 512 shots, i.e., 512 repeated measurements for a given fixed number of Grover iterations, in order to build sufficient statistical power, which is within the standard range in quantum computing simulation environments. We varied the number of Grover iterations, i.e., phase-flip marking of network states in \(S_n\), followed by an amplitude amplification diffusion step, from \(0\) to \(20\). Note that in our code, all values were encoded as decimal representations of the 7-bit binary numbers in \( \{0,1\}^7 \). For example, the state represented by the binary state \( \ket{0000101} \) is the number \( 5 \) in decimal form, i.e. $5_{10} = 0000101_{2}$. There is an (expected) relationship between \(|S_n|\) and the number of peaks in the frequency histogram: Over the range of iterations of the algorithm, as the number of states in \(S_n\), i.e. the size of \(|S_n|\), increases, the number of peaks, or frequency of oscillations, of the histogram increases monotonically over our test range of \(|S_n| \leq 50\) (Figure \ref{fig:grover}D). 

For a single marked state, after approximately \(\sqrt{N}\) Grover iterations among \(N\) possible states, the probability of measuring the marked state becomes high. However, given how we structured the network dynamics question, we needed to mark and simultaneously amplify several, possibly many, states in \(S_n\subset \{0,1\}^n\). In this scenario, the amplitude-amplification behavior of the algorithm and its measurement outcomes are different from the standard single-marked case, as our results reproduce from the known literature \cite{Boyer1996,Brassard2000}. 

There are two considerations to consider when \(|S_n| > 1 \). First, when \(S_n\) contains \(|S_n|\) elements, the amplitude amplification is necessarily distributed \emph{across all} the marked states in \(S_n\). The per-state amplitude growth would be expected to be significantly less than in a single-state case. In practice, each Grover iteration only partially shifts amplitude from the unmarked states into \emph{all} marked states. This means that with multiple marked states, the peak probability has to be shared among all \(|S_n|\) elements. Second, we expect to see an increased frequency of amplitude oscillations as \(|S_n|\) increases. This is what we indeed observed in numerical experiments of the Grover sub-circuit isolated from the rest of the analysis workflow, which is consistent with the published literature. As the size of the marked set increases, it produces a larger effective rotation angle per iteration during the diffusion step, causing the success amplitude in the marked subspace to rise and fall more frequently. Analyses of Grover to marking multiple states have shown that for some number of states \( k >1 \), the algorithm’s success probability oscillates at a rate proportional to \(\sqrt{k} \), with multiple peaks occurring for iterations past the first maximum.

So if \(|S_n|\) is a significant fraction of the total state space \( |\mathbb{S}_n| \), getting to a large individual state probability would require a more specialized version of the Grover algorithm. However, this is not necessary for our workflow here.

\subsubsection{Implications of Grover oscillations for our Deutsch--Jozsa approach and full workflow}
\label{subsec:grover-oscillations}

The oscillations that result when we use Grover to mark multiple states in \(S_n\) do \emph{not} affect our unique use of Deutsch-Jozsa in our problem construction and analysis workflow. While the output of the Grover step in the multi-solution condition causes amplitude oscillations among the marked states (Figure \ref{fig:grover}), the logic of our Deutsch--Jozsa runs — namely, distinguishing \lq all \(0\)' versus \lq all \(1\)' versus a mixed output that we can interpret as the potential for continued dynamics — does \emph{not} require each individual state in \(S_n\) to achieve a high probability of being measured. It only requires that they be marked in order to construct \(S_n\ \in \mathbb{S}_n\) so that we can evaluate \(\mathcal{T}(s_i)\) to construct the first part of \(U_f\), and alternatively assign constant values of \(0\) and \(1\) to its complement \(S_n^c\ \in \mathbb{S}_n\) for non-realized values for the second part of \(U_f\) before the Deutsch-Josza sub-circuit. 

Grover oscillations do not break the required marking of target states in \(S_n\). Even if the probability amplitudes among many marked states vary at different iteration counts, it remains true that \emph{all} states in \(S_n\) receive the phase-flip each time the oracle is applied. Hence, even with multiple solutions and increasing complicated oscillation patterns, the oracle is still \emph{recognizing} the correct subset \(S_n\). In other words, the circuit does mark those states, regardless of how the amplitude distribution evolves over repeated iterations.

Our two-pass use of the Deutsch--Jozsa sub-circuit, i.e., one run with complement \(=0\) and a second run with complement \(=1\), simply tests whether the function is globally constant or not. It does not rely on or require measuring a particular single basis state with high probability. Instead, it is the interpretation of the final measurement pattern that tells us whether the function is \lq all \(0\)' or \lq all \(1\)' or neither, according to the constructive/destructive interference in the final Hadamard stage. 

Yet, we still do gain quantum speedup from the superposition in the Grover step. Using Grover lets us identify membership in \(S_n\) without needing to check each of the \(2^n\) potential states individually (or \(\{0,1\}^7 = 128 \) states in our specific test example here). Quantum amplitude amplification \emph{is} happening, albeit spread among multiple solutions.

\subsection{Threshold Comparison Binary Subtraction Sub-Circuit}
\label{sec:threshold}

The threshold sub-circuit compares a node’s measured (or encoded) running summation value \(\Sigma_r\) (encoded by its corresponding $s_i$) at the observation time $T_o$ against a chosen threshold value $\Sigma_{T}$. It determines whether a state $s_i \in S_n$ exceeds the assigned constant value of $\Sigma_{T}$. Concretely, we represent values in $n$-bit binary, 7-bit in our test example and simulations here, and we need to determine if $\Sigma_{r} \ge \Sigma_{T}$. Once the code determines that the \(\Sigma_r\) for a given node summation exceeds the threshold, it declares that node as having \lq fired' and evaluates \(\mathcal{T}(s_i)\) to the binary output $=1$. Otherwise, it evaluates to $=0$.

\paragraph{Quantum circuit implementation.}
We store $\Sigma_{T}$ as an $n$-bit value, $\tau_{n-1}\dots\tau_{1}\tau_{0}$, i.e. 7-bit in our simulations. A binary \lq greater or equal' check is done bit-by-bit from the most significant to least significant position. If the first bit of a given $s_i$ with value $\Sigma_r$ exceeds that of $\Sigma_T$, we know $\Sigma_r \ge \Sigma_{T}$ without needing to check the remaining bits (\emph{c.f} section \ref{sec:bin_check} for details). This can be encoded as a multi-controlled $\mathbb{Z}$-gate sequence:

\begin{enumerate}[i.]
    \item \textbf{Initialize comparison qubits.} Load $\Sigma_{T}$ (in classical read-only form or) stored in dedicated qubits set to $\tau_{n-1}\dots\tau_{1}\tau_{0}$.
    \item \textbf{Iterate from most significant to least significant bit.} 
      For bit position $i$ (starting at $n\!-\!1$), compare $\nu_i$ (the bit of $\Sigma_r$ for a given $s_i$) to $\tau_i$ (the threshold bit in $\Sigma_T$). 
      \begin{enumerate}[a.]
        \item If $\nu_i = 1$ and $\tau_i = 0$, we immediately conclude $\nu \ge \Sigma_{T}$. 
        \item If $\nu_i = 0$ and $\tau_i = 1$, we conclude $\nu < \Sigma_{T}$.
        \item Otherwise, continue to the next bit.
      \end{enumerate}
    \item \textbf{Flip an ancilla for $s_i$ when $\Sigma_r \ge \Sigma_{T}$.} 
    After the logic identifies $\Sigma_r \ge \Sigma_{T}$, it sets an ancilla qubit for the marked state $s_i \in S_n$ to $\ket{1}$ (phase-flips it) to indicate that the threshold condition is satisfied. If $\Sigma_r < \Sigma_{T}$, the ancilla remains $\ket{0}$ (unflipped).
\end{enumerate}

In practice, our code arranges a multi-controlled operation in that for each $s_i \in \mathbb{S}_n = \{0,1\}^7$ (or $\in \{0,\dots,2^n-1\}$ in the general case) \emph{if} $s_i \ge \Sigma_{T}$ and $s_i \in \mathbb{V}_n$, we phase-flip the ancilla. More sophisticated, this can be constructed as a specialized \emph{quantum comparator} circuit, typically using $O(n)$ Toffoli and CNOT gates to implement a bitwise compare-and-flip.

\paragraph{Thresholding as a dictionary logic implementation.}
We also implemented the thresholding operation as a loop over all $s_i \in \mathbb{S}_n = \{0,1\}^7$ by evaluating
\[
  \text{if } (s_i \in S_n) \wedge (\Sigma_r \text{ (corresponding to } s_i \text{) } \ge \Sigma_{T}), \text{ then flip phase.}
\]
This is functionally a conceptual lookup approach. This is, of course, suboptimal from a minimal-gate-count perspective, but it makes the logic clear. The code effectively evaluates \lq For states $s_i$ that are acutally realized \emph{and} exceed the threshold, do a multi-controlled-$Z$ phase flip on an ancilla qubit.' This sub-circuit then attaches to the rest of the algorithm (see below), so the ancilla’s flipped state indicates $\mathcal{T}(s_i)=1$ if $\Sigma_r \ge \Sigma_{T}$, and $\mathcal{T}(s_i)=0$ otherwise, as required.

\subsection{Deutsch--Jozsa Sub-Circuit}
\label{sec:dj_subcircuit}

The final sub-circuit in our algorithmic workflow is the two-run Deutsch-Jozsa test. Once each $s_i \in S_n$ is evaluated and assigned $0$ or $1$, based on the thresholding sub-circuit, we need to assign fixed values of 0 and 1 to all $s_i \in S_n^c$, i.e. to the network states that never occured, in sequential Deutsch-Jozsa runs. This allows us to determine if $U_f$ for the \emph{entire} input range $\mathbb{S}_n = \{0,1\}^7$ is \emph{constant} or \emph{not constant}. This is necessary because Deutsch-Jozsa requires evaluating a function $f:\{0,1\}^n \to \{0,1\}$ (\emph{c.f.} section \ref{sec:DJ} above.)

More formally:
\begin{enumerate}[i.]
    \item If $s_i \in S_n^c$, then $\mathcal{T}(s_i)$ is forced to $0$ in the first run, and forced to $1$ in the second run.
    \item If $s_i \in S_n$, then $\mathcal{T}(s_i)$ is set by the thresholding operation above. This is functionally the evaluation of equation \ref{eq:fSn}.
\end{enumerate}

In our code, we define an oracle that, for each $s_i$, flips an ancilla phase if $\mathcal{T}(s_i)=1$, as discussed in the last section. When we then apply the standard Deutsch-Jozsa transformations, i.e., Hadamards before and after the oracle, an outcome of $\ket{0}^{\otimes n}$ in the final measurement still indicates an evaluated output function of constant = 1, whereas $\ket{1}^{\otimes n}$ still indicates a constant = 0, if and only if \emph{all} $ s_i \in S_n$ are 1 or 0. Any comparative deviation from a standard Deutsch-Jozsa output indicates a mixed network state. In practice, given our partial assignment and the small size (mixed value) $S_n$ we used in our numerical simulations, a large amplitude in $\ket{0}^{\otimes n}$ or $\ket{1}^{\otimes n}$ can arise even when $U_f$ is \emph{not} purely constant. This happens because destructive or constructive interference occurs for all states outside $S_n$ given the forced fixed nature of $S_n^c$, which artificially but correctly amplifies the dominant computational basis state. Nonetheless, we are able to measure and interpret mixed-state results, as required by the problem setup and algorithm, thereby allowing us to answer the question we pose in this paper (see Boxes 1. and 2. above). In other words, the \emph{combination} of threshold-based marking and the \emph{two-run} Deutsch-Jozsa test allows us to determine whether $S_n$ is purely $0$ or purely $1$ or is a mixture of states. The encoded logic allows us to interpret the network’s actual states $S_n$ as quiescent, epileptic, or having the potential to sustain inherent activity, as required.

\subsection{Numerical Results and Interpretation}
In this last section, we illustrate end-to-end three specific examples and their results, simulating, in turn, a mixed-state dynamics network capable of continuing inherent activity, a quiescent network, and an epileptic network. These examples are representative of all other runs we did for each condition. As introduced above, we assumed a (classical) network consisting of six nodes with evolving dynamics determined by the fractional summation model outlined in section \ref{sec:fracsum} up to some arbitrary observation time $T_o$. At $T_o$, each of the six nodes $v_j$ was manually assigned (i.e., arbitrarily simulated) a running summation value $\Sigma_r$ represented as a 7-bit value (\textit{c.f.} section \ref{sec:setmem} above). As such, there would be at most $2^7 = 128$ possible $\Sigma_r$. All three examples assumed the same six states for a toy six-node network, with $\mathbb{V}_n = \{20, 34, 42, 67, 81, 97\}$, i.e., $n=6$. We arbitrarily fixed the number of Grover iterations for all runs to 100 to yield $S_n$. Recall that in our code and outputs all 7-bit values were represented in standard decimal format.

\begin{figure} 
  \includegraphics[width=6.4in]{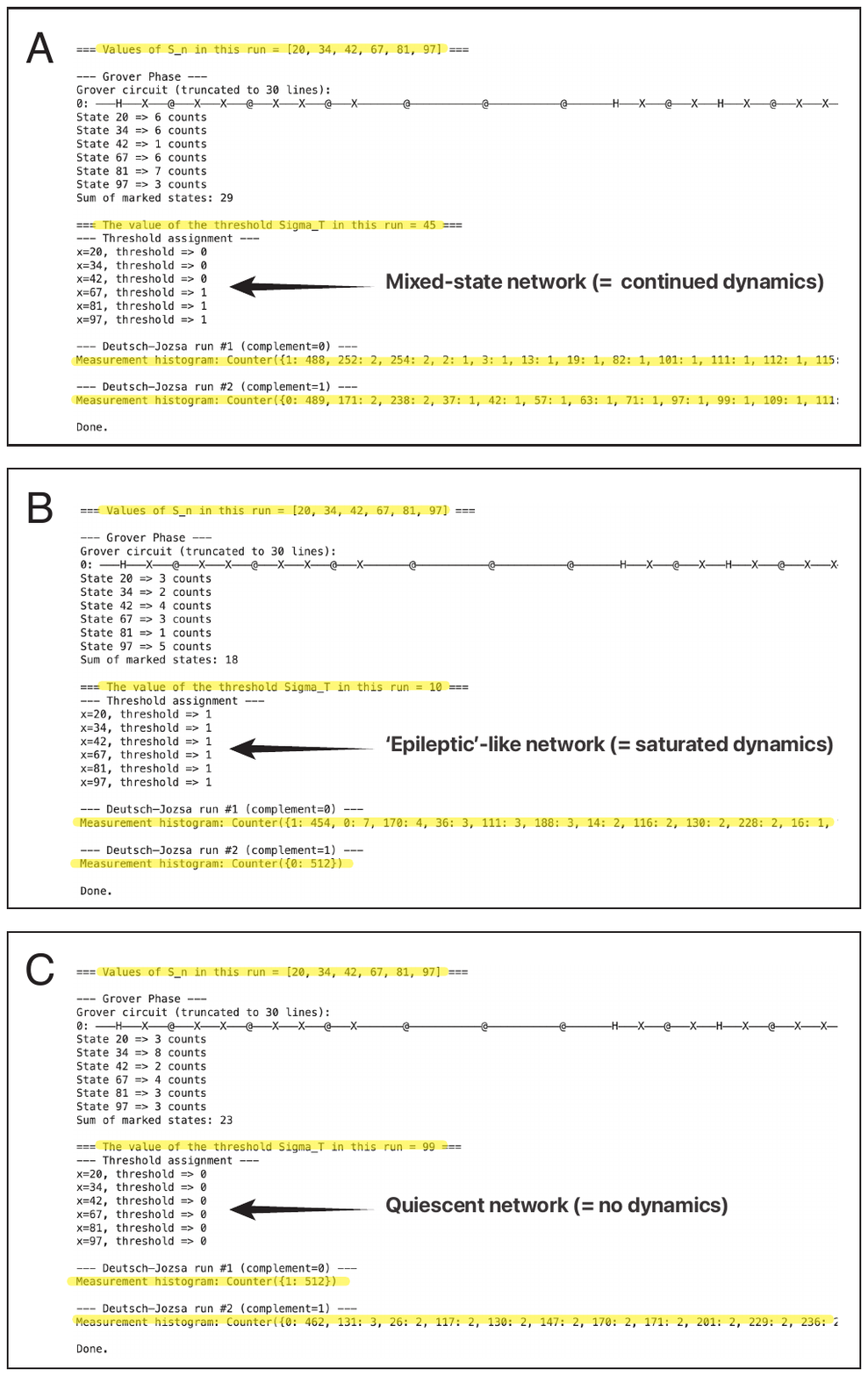}
  \caption{Summary outputs (screen shots) for the chained sub-circuits and workflow shown in Figure \ref{fig:fullcirc} for three different representative simulation conditions corresponding to mixed-state (\textbf{Panel A.}), \lq epileptic-like' (\textbf{Panel B.}), and quiescent (\textbf{Panel C.}) network dynamics. The data shows, respectively, for each panel, the marked states $S_n$, the threshold value $\Sigma_T$, the resulting network (dynamic) activation pattern, and the frequency histogram for each of the two runs of the Deutsch-Jozsa sub-circuit for each of the assignments $S_n^c \to 0$ and $S_n^c \to 1$.}
  \label{fig:DJ}
\end{figure}

Figure \ref{fig:DJ} shows the final output of the chained sub-circuits and workflow (\emph{c.f.} Figure \ref{fig:fullcirc} above) for each of the three examples. 

\subsubsection{Mixed-state network dynamics encoded by non-constant Deutsch-Jozsa measurements}
In \textbf{Panel A.} we assumed a threshold value $\Sigma_T = 45$, which was intentionally chosen so that upon thresholding $S_n$ it produced a set of mixed-states consisting of three \lq 0's (i.e. non-firing nodes) and three \lq 1' (i.e. active or firing nodes). 

This assigned an output to each $s_i \in S_n$ given by
\[
\mathcal{T}(s_i) = s_i \;\mapsto\;
\begin{cases}
1, & \text{if } s_i \geq 45,\\
0, & \text{if } s_i < 45.
\end{cases}
\]
Mapping
\[
\begin{aligned}
&20, \,34, \,42 \;\rightarrow\;0, \\
&67,\,81,\,97 \;\rightarrow\;1.
\end{aligned}
\]
The post-thresholded and updated $S_n$ then got passed to the Deutsch-Jozsa sub-circuit. For all $s_i \notin S_n$, the circuit assigned either $0$ (in the first run) or $1$ (in the second run). 

For the \textbf{first Deutsch-Jozsa run} (\boldmath{$S_n^c \to 0$}), this produced a combined two-part $U_f$ consisting of
\[
U_f \;=\;
\begin{cases}
1, & \text{if } s_i \in \{67,\,81,\,97\},\\
0, & \text{if } s_i = \{20,\,34,\,42\},\\
0, & \text{if } s_i \notin S_n \equiv s_i \in S_n^c
\end{cases}
\]
for the full range of values $[0,127]$, i.e. $\mathbb{S}_n = \{0,1\}^7$.

For the \textbf{second Deutsch-Jozsa run} (\boldmath{$S_n^c \to 1$}), the combined two-part $U_f$ was
\[
U_f \;=\;
\begin{cases}
1, & \text{if } s_i \in \{67,\,81,\,97\},\\
0, & \text{if } s_i = \{20,\,34,\,42\},\\
1, & \text{if } s_i \notin S_n \equiv s_i \in S_n^c
\end{cases}
\]

The resulting frequency histogram for the first run consisted of a mixture of states dominated by \texttt{1:488 counts}. The second run also produced a mixture of states, dominated by \texttt{0:489 counts}. In both cases, the dominant basis given the forced assignments $S_n^c \to 0$ in the first run and $S_n^c \to 1$ are as expected, along with a mixture of other states. (See the discussion in the preceding section \ref{sec:dj_subcircuit}.) Assigning $S_n^c$ to $0$ or $1$ in each pass extended $\mathcal{T}(s_i)$ over all $128$ states. We were able to show from the thresholding logic that $\mathcal{T}(S_n)$ was \emph{not} globally constant: Some realized network dynamic states were $0$, and others were $1$, which agree with the construction of the toy example. From this, we can interpret the output of the circuit workflow that the network dynamics has the potential to continue sustained activity rather than being purely quiescent ($0$) or epileptic-like ($1$).

\subsubsection{\lq Epileptic'-like network dynamics encoded by constant Deutsch-Josa measurements}
In \textbf{Panel B. of Figure \ref{fig:DJ}} we repeated the simulation but re-assigned the threshold value to $\Sigma_T =  10$. This resulted in an output from the thresholding circuit of 1 for all $s_i \in S_n$, emulating a network with all nodes active (firing) at the observation time $T_o$. 

When $S_n$ was then used as the input into each of the two Deutsch-Jozsa runs, it produced 

\[
U_f \;=\;
\begin{cases}
1, & \text{if } s_i \in \{20,\,34,\,42\,67,\,81,\,97\},\\
0, & \text{if } s_i \notin S_n \equiv s_i \in S_n^c
\end{cases}
\]
for the \textbf{first Deutsch-Jozsa run} (\boldmath{$S_n^c \to 0$}) and 
\[
U_f \;=\;
\begin{cases}
1, & \text{if } s_i \in \{20,\,34,\,42\,67,\,81,\,97\},\\
1, & \text{if } s_i \notin S_n \equiv s_i \in S_n^c
\end{cases}
\]
for the \textbf{second Deutsch-Jozsa run} (\boldmath{$S_n^c \to 1$}).

The workflow and two runs of the Deutsch-Jozsa sub-circuit again behaved as expected, with the interpretation of the final measurements correctly identifying \lq epileptic'-like network dynamics. While the output of the first run was mixed, the output of the second run resulted in a frequency histogram \texttt{0:512 counts}, i.e., no mixed states, as expected.

\subsubsection{Quiescent network dynamics encoded by constant Deutsch-Josa measurements}
In the last experiment we did, with outputs shown in \textbf{Panel C. of Figure \ref{fig:DJ}}, we again repeated the simulation keeping $\mathbb{V}_n$ the same,  but this time re-assigned the threshold value to $\Sigma_T =  99$. In this scenario, the output from the thresholding circuit resulted in 0 for all $s_i \in S_n$, emulating a network with all nodes not firing in a quiescent state at the observation time $T_o$. 

In this case, $S_n$ and therefore $U_f$ was reversed in contrast to the \lq epileptic'-like network:
For the \textbf{first Deutsch-Jozsa run} (\boldmath{$S_n^c \to 0$}) 
\[
U_f \;=\;
\begin{cases}
0, & \text{if } s_i \in \{20,\,34,\,42\,67,\,81,\,97\},\\
0, & \text{if } s_i \notin S_n \equiv s_i \in S_n^c
\end{cases}
\]
While for the \textbf{second Deutsch-Jozsa run} (\boldmath{$S_n^c \to 1$})
\[
U_f \;=\;
\begin{cases}
0, & \text{if } s_i \in \{20,\,34,\,42\,67,\,81,\,97\},\\
1, & \text{if } s_i \notin S_n \equiv s_i \in S_n^c
\end{cases}
\]

Again, the workflow and two runs of the Deutsch-Jozsa sub-circuit produced an interpretable output that correctly identified quiescent network dynamics. The output of the first run was constant with a frequency histogram \texttt{1:512 counts}, while the output of the second run produced a mixed histogram.

This toy example is representative of all numerical simulations we attempted. We observed no anomalous deviations from the expected behavior of the quantum circuits and workflow for any variations in the values of the workflow's variables, including specific decimal values of $\mathbb{V}_n$, the size of $\mathbb{V}_n$, i.e., $|\mathbb{V}_n| = |S_n|$ (up to 20), the number of Grover iterations (up to 100), or the threshold values of $\Sigma_T$.

\section{Discussion}

In this work, we introduced a particular class of problem related to the dynamics of large-scale networks relevant to neuroscience and machine learning. The problem we addressed asks whether a network can sustain inherent dynamic activity, cease functioning through quiescence, or saturate in an epileptic-like state. By carefully structuring the problem, we were able to take advantage of the unique properties of quantum computing, particularly the Deutsch–Jozsa and Grover algorithms, and provide an approach that solves this problem with significant computational efficiency compared to classical methods that scales with the size of the network.

Classically, evaluating the dynamics of large networks requires an exhaustive search through the network's state space, a computationally prohibitive task for very large networks. In contrast, the quantum approach we introduce here exploits superposition to evaluate the global state of the network in parallel, reducing the computational complexity by orders of magnitude. 

\subsection{Theoretical and Applied Implications to Biological and Artificial Neural Networks}
The importance of this work lies in both its theoretical and practical implications. For neuroscience, it provides a novel approach to model and understand the temporal and structural properties of neural networks. More broadly, it has potential applications for analyzing how brain dynamics relate to disorders associated with network disconnections, such as epilepsy, neurodevelopmental disorders such as Autism Spectrum Disorder, and neural cognitive and degenerative disorders. In machine learning and artificial intelligence, it is potentially relevant to constructing and optimizing artificial neural network models. For quantum computing, this work serves as a proof-of-concept for how existing quantum algorithms can be adapted and extended to solve applied, non-trivial interdisciplinary problems, motivating the continued exploration and expansion of quantum algorithms and computing.

Of particular relevance are the comments we expanded on in section \ref{sec:groveradvantage} above. An extension of the ideas we explore here could lead to a novel analog-digital neural dynamic model that more closely resembles how we know biological neurons compartmentalize processes and compute. As we briefly introduced, biological neurons are not just purely digital computational units but rather highly complex analog-digital computing elements individually that collectively contribute to network dynamics. An approach that leverages the inherent properties of the Bloch sphere or an evolving Hamiltonian before collapsing to a binary output could be a unique way to think about an analog-digital hybrid neural dynamic model. Such a model could conceivably have applicability to understanding how the biological brain uniquely represents and processes information while also forming the basis of new forms of quantum machine learning and artificial neural networks. In such a model, the evolution of the network may not be observable.

\subsection{Modeling the Evolution of Network Dynamics Using Quantum Computational Methods}
This work raises two broad areas in which quantum computing may contribute to the scientific study of neural networks. The first is large-scale simulations of neural dynamics across scales of spatial and temporal organization, bounded and informed by the known physiology (in the brain) and known models (for artificial intelligence and machine learning). We did not address this first question here. The second is carefully chosen and structured problems \textit{about} neural dynamics that use quantum algorithms carefully and in a clever way, an example being the work in this paper. These two research directions are necessarily complementary rather than mutually exclusive or distinct.

Sufficiently large-scale simulations would allow observing, experimenting, and iterating numerical experiments under a wide range of parameter and model conditions. This is important because if, as neuroscientists suspect, complex emergent cognitive properties are partly due to sufficiently large interactions among foundational physiological and biological components and processes across temporal and spatial scales of organization, the need to carry out large iterative simulations may be critical to understanding or even just observing the dynamics that give rise to emergent cognitive properties. Simulations of this kind would support understanding emergent effects that depend on the scale of the computational space.  

But open-ended large-scale simulations alone will not be sufficient for understanding how the brain, or artificial intelligence for that matter, works. In effect, open-ended large-scale simulations in isolation led to the significant challenges and missed targets faced by the highly publicized and hugely funded Blue Brain Project \cite{nature2020}. Arriving at a deep understanding necessitates carefully chosen and defined problems \textit{about} the neural network dynamics being simulated. This is critical. Observing neural dynamics in action — for example, the firing patterns of large numbers of neurons — in isolation and without context alone cannot reveal the underlying algorithms operating on those dynamics or why they exist. It is no different than attempting to understand the brain from a systems perspective by studying a single participating protein or ion channel in isolation. Quantum computing potentially has a unique role in deciphering and understanding the contributions of functional network dynamics and behaviors on carefully formulated and constructed \textit{interpretable} questions and problems, similar to what we do in this paper. 

As also discussed above in section \ref{sec:groveradvantage}, there may also be scenarios where, instead of a network model being simulated, a network evolves naturally according to its physical makeup, i.e., activity in the brain or internal activation patterns in artificial neural networks. In this scenario, only then are the dynamics measured or observed at some arbitrary time $T_o$ to probe its behavior. In this case, as we showed here, a quantum computational approach to a well-structured question about the dynamics could be the only way to answer the question. This could have real-world implications. The dynamics of the network may be inaccessible or uninterpretable before or beyond $T_o$. For example, in biological neural networks, experimental limitations, such as noise, finite measurement windows, or ethical considerations, may constrain observations to specific time windows. 

Our quantum framework and approach focus on evaluating the global state of the network at $T_o$ all at once to infer its dynamic state and potential. Unlike classical methods, which necessitate the individual evaluation of each node and which quickly becomes computationally intractable, the quantum approach leverages superposition to assess the entire network state simultaneously. This enables insights into the network's ability to sustain activity or transition between states without granular interrogation of the individual nodes that make up the network. The kind of framework we developed here provides a novel quantum computational lens for studying network dynamics.

\section*{Acknowledgments}
The author would like to thank Robert Treuhaft, with whom he had some of the earliest conversations related to quantum computing and the brain. And to Eric Traut and Rita J. King for many insightful discussions and feedback.

\bibliographystyle{naturemag}
\bibliography{refs}

\end{document}